\documentclass[sigconf,nonacm]{acmart}
\usepackage{graphicx}

\usepackage{algorithm} 
\usepackage{algpseudocode}
\usepackage{cleveref}
\newtheorem{definition}{Definition}
\newtheorem{lemma}{Lemma}
\newtheorem{corollary}{Corollary}
\newtheorem{theorem}{Theorem}
\newtheorem{remark}{Remark}

\begin{document}

\title{Private Vulnerability Testing for User Oversharing}

\date{}

\author{Alexander Boyd}
\orcid{0009-0007-3322-9556}
\affiliation{%
  \institution{University of Surrey}
  \city{Guildford}
  \country{United Kingdom}}
\email{ab01353@surrey.ac.uk}

\author{Daniel Gardham}
\affiliation{%
  \institution{University of Surrey}
  \city{Guildford}
  \country{United Kingdom}}
\email{daniel.gardham@surrey.ac.uk}

\author{Nishanth Sastry}
\affiliation{%
  \institution{University of Surrey}
  \city{Guildford}
  \country{United Kingdom}}
\email{n.sastry@surrey.ac.uk}

\author{Suparna De}
\affiliation{%
  \institution{University of Surrey}
  \city{Guildford}
  \country{United Kingdom}}
\email{s.de@surrey.ac.uk}

\begin{abstract}
Users can inadvertently increase their vulnerability to privacy attacks by oversharing personal information. In particular, the combination of individually innocuous attributes can place a user in a sufficiently small population to enable linkage or re-identification attacks. We present a privacy-preserving mechanism that allows users to determine whether the information they disclose places them below a predefined population threshold, without revealing their query or the data of other users. Our setting presents several practical challenges: users contributing data may be offline when queries are made, queries may select arbitrary subsets of attributes, and neither server should learn the queried population or its size. We address these challenges using secret-shared data and queries, private equality testing, and secure threshold computation. Our protocol provides computational privacy against a malicious client and semi-honest, non-colluding servers, while supporting arbitrary queries and dynamic database updates. We define reactive $f$-privacy for a dynamically evolving database and prove that our construction satisfies it, with the sequence of permitted threshold outputs defining the information available to the adversary across an execution. We optimise the protocol using oblivious-transfer extensions and pre-computation, achieving sub-second online query times for 96-bit queries on datasets of up to one million records.  
\end{abstract}

\maketitle

\section{Introduction}

Despite a majority of Internet users having concerns about privacy online, many struggle to translate this into actionable privacy protection measures~\cite{story2021}. An important reason for this is that users may not understand the extent to which they may be oversharing or the dangers involved in such \cite{oversharing, oversharing2}. Individuals disclose information about themselves over time (across multiple posts) and often across multiple online and offline sources. For example, a user may publicly reveal their approximate
age, location, employer, and university through social media and professional profiles. While each attribute may be relatively non-identifying in isolation, their intersection may correspond to
only a small number of individuals, making the user vulnerable to linkage with a sensitive dataset. Similar risks arise in health and mobility data, where combinations of demographic, geographic, and behavioural attributes can be sufficient to distinguish individuals. 

By linking all the attributes shared, adversaries might be able to identify individuals. The power of linkage attacks has been demonstrated through examples such as Narayanan and Shmatikov's~\cite{narayanan2008} classic de-anonymisation, using IMDb, of a Netflix dataset with 500,000 users' ratings in order to learn sensitive information, and the reidentification of taxi trips by Pandurangan~\cite{taxilogs}---a huge dataset of taxi trips granted from a Freedom of Information request, which contained hashed personally identifiable information (PII). 

Although linkage attacks have often targeted high-profile individuals, people belonging to small or under-represented populations may also face heightened risk, since relatively few additional attributes may be sufficient to distinguish them from the wider population. For example, if transgender individuals constitute a very small proportion of a population, they may be singled out in a database using only a few additional attributes. Such individuals may therefore need to consider carefully whether disclosing additional information could make them more identifiable. 

In this paper, we define the vulnerability of a user for a given database to be the size of the population determined by the publicly available information of that user. Thus, an attribute does not have an intrinsic level of linkage risk, which depends on what other information is available and on the distribution of those attributes in the population. 
The core functionality is simple: given a user's disclosed attributes, the system uses two non-colluding servers to privately count the number of users whose records match those attributes and returns only whether this count is below a predefined threshold $T$. Neither server learns the user's attribute values, the resulting count, nor any individual user's record. Unlike state-of-the-art private aggregation such as Prio~\cite{prio} and Prio+~\cite{prio+}, where the aggregate function is specified as part of the data-collection process, our setting allows users to privately query arbitrary subsets of attributes after data collection and determine the frequency of the corresponding joint attribute combination, even when the users contributing to the database are offline.

Moreover, the privacy guarantee must hold across the system's entire lifetime, rather than for each query considered in isolation. A querying client may adapt its next query to previous threshold outcomes, while insertions and updates may change the population against which subsequent queries are evaluated. Consequently, even if each individual execution reveals only its permitted output, the sequence of outputs may reveal additional information about the evolving dataset. We capture this stateful setting through \emph{reactive $f$-privacy}: the adversary's view across an adaptive sequence of queries and database operations must reveal no more than is permitted by the corresponding ideal functionality. Our definition therefore makes explicit the leakage arising from the allowed outputs and the evolution of the database.

We represent each user's record as a vector $v_i$ of encoded attribute values. Categorical attributes can be represented by enumerating their possible values (e.g., male, female, transgender), while numerical attributes, such as age, can be discretised into buckets, again creating an enumeration of possible values (e.g., 20--30, 30--40, etc.). A client's query consists of an attribute-selection vector $q$, indicating which attributes are being queried, together with the corresponding attribute values $t$. For example, a client may query whether the population matching their age, location, and sexuality is below a predefined threshold, without querying or revealing their other attributes. This representation allows users to select an arbitrary subset of attributes when making a query, while the underlying records remain fixed in the database.

Two non-colluding servers are required to maintain privacy. Organisationally, non-collusion might require that these servers are operated by two competing social media platforms or perhaps two charities, such as ACLU and EFF that care about privacy and civil liberties. Technically, non-colluding servers enable privacy of the users' attribute values by maintaining secret shares of users' records. Similarly, privacy of user queries is ensured by having the user's client secret-share its query between the servers. The servers privately perform equality tests between the query and each database record, aggregate the resulting matches, and securely determine whether the resulting count is below the threshold $T$.

Within the above framework, it is still challenging to perform a private intersection across arbitrary combinations of attributes without incurring an exponential blow up. We tackle this by using a secret-shared enumeration of the user's data and computing over it when needed. We utilise Beaver triples for secret multiplication that allows us to offload as much pre-computation as possible. This approach also allows us to very cheaply hide the query using a clever yet simple extension.

\paragraph{Our Contributions.}
We present a privacy-preserving vulnerability testing system that allows users to determine whether their disclosed attributes place them in a population below a predefined threshold, while keeping both the underlying user data and the querying user's attributes private. Our construction supports arbitrary attribute-subset queries, offline users, and a dynamically evolving database. We achieve practical online performance by moving computationally expensive operations to a pre-computation phase. Our contributions are as follows:
\begin{enumerate}
\item Private vulnerability testing. We provide a formal construction for privately determining whether a user's attributes identify a population below a predefined threshold, where the query may select an arbitrary subset of attributes and users contributing to the database need not be online when the query is issued. Our construction provides computational privacy against malicious clients and semi-honest, non-colluding servers.
\item Reactive privacy. We define reactive $f$-privacy for our stateful setting, capturing the security of an adversary's view across an arbitrary sequence of queries and database updates. We prove that our construction satisfies reactive $f$-privacy, with the permitted threshold outputs characterising the information revealed across the execution.
\item Extensions and optimisations. We show how the construction can support client opt-out and data updates, and formulate the equality-testing component as a replaceable interface, allowing alternative equality-testing techniques to be incorporated without modifying the remainder of the protocol. We additionally develop a more efficient variant that trades a modest amount of information leakage for improved performance.% and consider a resource-allocation mechanism to mitigate clients that submit deliberately expensive queries.
\item Implementation and evaluation. We implement the proposed protocols in C/C++\footnote{Available at: \url{https://anonymous.4open.science/r/privateintersect-26D7/}} and evaluate their performance on datasets of up to ten million records. We additionally implement an alternative equality-testing protocol and compare its performance with our baseline construction.
\end{enumerate}

\section{Related Work}
Privacy-preserving measurement (PPM) enables a party to perform measurements over client data without learning individual records. Several techniques have been proposed to achieve this, including differential privacy (DP), homomorphic encryption (FHE), and secret sharing. We consider these approaches in the context of privately computing the intersection count between a client's attributes and those of other \emph{offline} users. 

\paragraph{Differential Privacy.}
Differential privacy provides a well-established approach to protecting individuals when releasing aggregate statistics, including under repeated queries through privacy accounting and composition~\cite{dp-og,dp2}.
However, it introduces a privacy--accuracy trade-off through the addition of noise, whereas our application requires an exact threshold decision. This trade-off becomes particularly significant when many attribute
constraints must be evaluated, as the added noise can obscure whether the matching population lies above or below the threshold. Stronger, more recent proposals \cite{rappor,prochlo} mitigate this drawback, but in any case, the record of a client must inherently be similar to the ground truth.  Differential privacy also provides robustness against malicious clients, but this does not address our requirement for exact, post-hoc threshold queries over secret-shared data.

\paragraph{Fully Homomorphic Encryption.}
FHE may seem like a natural approach to achieve PPM, and there are indeed several works that take advantage of the inherent security offered by FHE \cite{FHE1, FHE2, FHE3}. However, despite possessing strong confidentiality properties, it has its own drawbacks. Supporting malicious clients would require additional verifiable computation techniques or mechanisms for enforcing valid inputs~\cite{vFHE}. Such mechanisms introduce further computational and communication overhead. Combined with the relatively high cost of FHE evaluation, this makes FHE unattractive under the resource and latency constraints of our system.

\paragraph{Sketching Protocols.}
Since we need only deal with counts, the advances in sketching protocols may seem alluring \cite{sketch1, sketch2, sketch3}. However, the challenge is not simply computing a count, but supporting counts over arbitrary intersections of attribute values. Consequently, representing every possible query as an element of the input domain leads to a combinatorial explosion as the number of attributes grows (i.e. the power set excluding overlap). Even with sub-linear space complexity of the codomain from sketch compression, we see an explosion as the system scales (for reasonable error values), and the approach becomes impractical. For example, an input domain of 64 bits has an upper bound of $2^{64} \approx 10^{19}$ outputs before sketch compression.

\paragraph{Secret Sharing.} 

Secret-shared aggregation systems such as Prio~\cite{prio} and Prio+~\cite{prio+} support private frequency estimation, but are designed around a measurement and aggregation function specified during data collection. Our setting instead requires clients to issue arbitrary intersection queries after the database has been populated, without requiring contributing clients to participate in subsequent queries. Encoding every possible attribute combination as a separate value, as in a one-hot representation, incurs combinatorial space complexity. We therefore store secret shares of the underlying attribute vectors and evaluate equality with the query at query time, allowing arbitrary post-hoc intersections without exposing the queried values.

\section{Technical Preliminaries} \label{sec:prelims}
This section introduces the cryptographic primitives used by our construction. In particular, we describe a two-party secret-sharing multiplication protocol, and show how it yields distributed AND and equality testing. Finally, we describe the Boolean-to-arithmetic (B2A) conversion used to aggregate equality-test outputs. 
\paragraph*{Notation and Assumptions}

We define various mathematical notation used throughout this paper. $\mathbb{F}^L_m$ is the $L$-dimensional finite field of size $m$. Bitstrings (bitvector or vector similarly) are defined from the domain $\mathbb{F}^L_2$. Logical operators are taken as their bitwise counterparts when invoked on a vector. $[x]^i$ denotes the $i^\text{th}$ Boolean-share of variable $x$, such that $\bigoplus_i [x]^i = x$. We additionally define the function AND$(\cdot)$ which takes a bitvector $v = (v_1 \dots v_L)$ and returns $\prod_{j=1}^L v_j$. The number of parties assumed for a Multiparty Computation (MPC) is 2 unless otherwise stated. To \emph{publish}, in the context of MPC, means to inform all other parties of the corresponding message.

\paragraph*{Dataset Format.}
A dataset or database is structured by an ordered set of $k$ attributes $\mathcal{A} = (\mathit{A}_1,  \dots , \mathit{A}_k)$. Each attribute has an associated finite domain of values, denoted $\mathcal{D}_{\mathit{A}_i}$ for attribute $\mathit{A}_i$. A record $\mathcal{R}_i$ is then a tuple of values where each value comes from each attribute exactly once, $\mathcal{R}_i = (d_{i1},  \dots , d_{ik})$, $d_{ij} \in \mathcal{D}_{\mathit{A}_i}$ for all $1 \leq j \leq k$. A dataset $\mathfrak{D}$ can naturally be viewed as a set of records $\mathfrak{D} = (\mathcal{R}_1,  \dots , \mathcal{R}_n)$. 

The bitlength of an attribute $\mathit{A}_i$ is then denoted $\mathcal{L}_{\mathit{A}_i} = $\newline$\lceil \text{log}_2(|\mathcal{D}_{\mathit{A}_i}| + 1)\rceil$ and the bitlength of a full record is $\ell = \sum_{i=1}^k{\mathcal{L}_{\mathit{A}_i}}$. The structure of a dataset is defined with the tuple \newline $\sigma = (\mathcal{A}, k, \ell, (\mathcal{D}_{\mathit{A}_1},\ldots,\mathcal{D}_{\mathit{A}_k}), (\mathcal{L}_{\mathit{A}_1}, \dots ,\mathcal{L}_{\mathit{A}_k}))$.

\paragraph*{Secret-sharing.}

%While DP achieves privacy through random noise and FHE through encryption, 
Secret-sharing works by decomposing a secret $x \in \mathbb{F}^m$ into $n$ \emph{shares}, such that $x = F([x]^1,  \dots , F[x]^n)$, for an arbitrary function $F$. Notably, we consider the \emph{Boolean secret sharing} scheme, which takes $F$ as bitwise XOR (exclusive OR) and secrets in $\mathbb{F}^L_2$, and the \emph{arithmetic secret sharing} scheme, which takes $F$ as field addition and secrets in $\mathbb{F}_L$. If we give each party $i \in (1 \leq i \leq n)$ the share $[x]^i$, then it is clear to see that these schemes are perfectly secure, in the sense that if a malicious party were to recover all but one of all shares, they would gain no information about the actual secret $x$.

\paragraph*{Arithmetic for Secret Shares.}
With addition and multiplication, it is possible to define a large class of functions. Addition of shares with a constant or another share is straightforward and requires no communication. To calculate the addition of two secrets $z = x + y$, each server locally computes $[z]^i = [x]^i + [y]^i$. 

Multiplication by a constant is also easy and similarly only requires local multiplication with the constant. Multiplication between secrets, however, is less trivial and requires communication. Although there are several ways to compute multiplication of secrets, for this paper we use the approach used by Beaver \cite{beaver}. In this approach, the multiplication is pre-computed using a random collaboratively computed multiplication, called a \emph{Beaver triple}. The triple is 'mixed' with the share of the secret which is to be multiplied, then published to the other parties. The parties then perform a computation that multiplies the secrets while removing the parts of the Beaver triple, as we shall now describe in detail.

\paragraph*{Generation of Beaver Triples.}
A Beaver triple is of the form $(a, b, c) \in \mathbb{F}^3$, where $c = ab$. At the end of the generation, party $i$ should have $([a]^i, [b]^i, [c]^i)$. A triple in $\mathbb{F}^3_L$ can be generated via \emph{oblivious linear evaluation}, which uses 2log$_2 L$ instances of \emph{oblivious transfer} (OT). For the Boolean field ($L=2$) used throughout this paper, we need only use 2 OTs. The exact protocol used is detailed in \Cref{triples}.

\paragraph*{Evaluation of Multiplications.} \label{distmulti}

In order to evaluate the multiplication of two secrets $z = xy$ and a Beaver triple $(a, b, c)$, each party begins by computing intermediate values $[d]^i = [x]^i - [a]^i$ and $[e]^i = [y]^i - [b]^i$. Each party then \emph{publishes} their share, such that they can reconstruct the real values of $d$ and $e$. Revealing these shares does not compromise the security of the secrets $x$ and $y$ since they are the addition of two secrets in a finite field, whose security argument is identical of that to an arithmetic secret sharing (they are indistinguishable from random). Finally, $[z]^i$ is calculated as $de + d [b]^1 + e [a]^1 + [c]^1$ for party 1 and $d [b]^2 + e [a]^2 + [c]^2$ for party 2. Proof of correctness is seen in \Cref{triples}. 

\paragraph*{B2A Conversion.}
Once in this form, the shares of the outputs from many equality tests cannot be aggregated, due to the possibility that the modulus be overflowed upon aggregation. In general, the shares need to be in a field of size at least $n + 1$, for $n$ outputs to be aggregated. Conveniently, the work of \cite{prio+} demonstrates how to utilise B2A conversions for data aggregation. 

The approach of B2A conversions follows exactly that of secret multiplication - an expensive pre-computation is performed on random inputs, using OTs to generate a correlated pair (a 'daBit', a double-authenticated bit) corresponding to a conversion from a Boolean field $\mathbb{F}_2$ to a larger one $\mathbb{F}_p$ - which is then used to perform another conversion later on, but with chosen inputs and much cheaper computation and communication. More concretely, the pre-computation requires just one OT in the offline phase and one bit of communication in the online phase. The protocols for generation and evaluation can be seen in \Cref{B2A}.

\paragraph*{Distributed AND and Private Equality Testing.}
A distributed AND function with arbitrary number of inputs can be trivially built using distributed multiplication. Consider that the AND function is equivalent to multiplication for Boolean inputs, then, the AND function can be computed over several inputs by sequentially multiplying all bits. 
An equality testing protocol can then be built using the distributed AND function. To test whether a shared bitstring $v = [v]^1 \oplus [v]^2$ is the vector of all 1s, one can run the distributed AND function over $v$ as previously defined. For a test vector $t$, this protocol can be extended to arbitrary equality testing by providing the bitmask $\neg t$ before testing, and it is easily verified that $t = v$ if and only if AND$(\neg t \oplus v) = 1$.

\section{System Model and Security Goals} 

We consider a system consisting of two non-colluding servers and a set of clients. Clients may independently submit records to the system or issue queries at any time. Each server stores a secret share of every submitted record, such that no individual server can reconstruct an honest client's data.

A key requirement of our system is that clients are \emph{offline} after submitting their data. In particular, a client who contributes a record is not required to remain online, respond to server requests, or participate in subsequent queries. Queries must therefore be evaluated entirely from the data already held by the servers. This requirement is important in our target setting, where clients may be
intermittently connected, unavailable when a query is issued, or unwilling to participate in repeated computations. Consequently, the cost of answering a query must not depend on the availability or computational resources of the clients whose data contributes to the result.

Let $\mathcal{D}=(R_1,\ldots,R_n)$ denote the current database. A query $Q=(q,t)$ specifies a subset of attributes $q$ and corresponding attribute values
$t$. Let $\mathsf{count}_\mathcal{D(Q)}$ denote the number of records in $\mathcal{D}$ whose values agree with $t$ on the attributes selected by $q$. The system evaluates the vulnerability predicate %\dan{propagate this terminology throughout}
\[
\mathsf{Vuln}_T(\mathcal{D},Q)=
\begin{cases}
1 & \text{if }\mathsf{count}_\mathcal{D}(Q)<T,\\
0 & \text{otherwise.}
\end{cases}
\]
The client learns only this Boolean result. In particular, neither the client nor either server learns the exact value of $\mathsf{count}_D(Q)$. Numerical attributes can be represented using publicly defined buckets, allowing the representation to trade precision for both computational efficiency and privacy. Categorical attributes can be represented directly using the same enumeration mechanism.

The threshold $T$ is a public system parameter selected by the system operator. It represents the minimum population size considered sufficiently large under the system's privacy policy. Clients cannot select an arbitrary threshold for individual queries, since doing so would permit an adversary to recover population sizes by repeatedly querying with different thresholds. A system may instead expose a fixed
set of predefined thresholds, if required. 
\noindent Our system is designed to provide the following properties:

\begin{description} 
\item{Correctness.} \label{reqs}
For an honestly generated database and honestly generated query, the client receives the correct value of $\mathsf{Vuln}_T(D,Q)$, except with negligible probability.
\item{Robustness.}
A malicious client should be unable to affect the protocol or the records of other clients beyond the effect achievable by submitting an arbitrary valid record on its own behalf.
\item{Privacy.} An adversary controlling one server should learn no information about the records of honest clients or the values of a client's query beyond the public parameters and the outputs permitted by the security definition. In particular, the servers should not learn either the queried attribute values or the exact matching population size. Since the database may evolve and queries may be issued repeatedly, we require this privacy guarantee to hold across an arbitrary sequence of queries and database updates. We formalise this stateful security requirement as reactive
$f$-privacy in \Cref{sec-proof}.
\end{description}

\paragraph{Threat Model.}

We consider an adversary that may corrupt one of the two servers and an arbitrary number of clients. The corrupted server is \emph{semi-honest}: it follows the prescribed protocol but may inspect its complete internal state and execute any additional efficient computation on its view. The two servers are assumed not to collude.

Corrupted clients may deviate arbitrarily from the protocol, including submitting malformed or adversarially chosen records and issuing arbitrary queries. We require that such clients cannot compromise the privacy of honest clients or affect the database beyond their own permitted contributions.

We assume authenticated, confidential communication between all parties. We do not consider collusion between the two servers or malicious deviation by either server. Consequently, correctness and robustness are guaranteed against arbitrary malicious clients provided that both servers follow the protocol, while privacy is guaranteed
against an adversary controlling at most one server together with any number of clients.

Formal security definitions, including $f$-privacy and its reactive extension, are given in \Cref{sec-proof}.

\section{System Design}

Our construction is based on a simple observation: the desired query can be expressed as a private intersection between the client's query and the secret-shared database. Each server holds only a share of every client record
and of the query, allowing the servers to determine whether each record matches the query without reconstructing either input.

At a high level, the protocol proceeds in three stages. First, the servers perform a private equality test between the query and each database record. Second, the resulting equality bits are converted and aggregated to obtain a secret-shared count of matching records. Finally, the servers privately compare this shared count with the public threshold $T$, revealing only the resulting Boolean vulnerability decision.

A key design requirement is that clients are offline after submitting their records. Consequently, all computation required to answer a query is performed by the two servers using the data already stored in their secret-shared databases. Unlike interactive private aggregation protocols in which clients participate
in each computation, our construction separates data submission from query execution. A client participates only when submitting, updating, or removing its own record. Once the record has been secret-shared between the servers, subsequent queries can be evaluated without any interaction with that client.

The equality tests produce a secret-shared count rather than a public count. Reconstructing this count would violate the intended privacy guarantee: an adversary could use repeated queries to identify individual records. For example, if an adversary knows that a particular individual belongs to the database, a query that fixes all but one attribute could reveal the unknown attribute if the resulting population has size one. We therefore require the threshold comparison itself to be performed without reconstructing the count. The servers use a garbled circuit (GC) to evaluate whether the secret-shared count satisfies the vulnerability predicate, revealing only a masked Boolean value of the vulnerability predicate. The mask prevents either server from learning the comparison result directly. We therefore arrange the output-sharing mechanism so that the result is delivered to the client, who is entitled to learn the vulnerability decision, while neither server obtains the result individually.
The equality tests produce a secret-shared count rather than a public count. Reconstructing this count would violate the intended privacy guarantee: an adversary could use repeated queries to identify individual records. For example, if an adversary knows that a particular individual belongs to the database, a query that fixes all but one attribute could reveal the unknown attribute if the resulting population has size one. We therefore require the threshold comparison itself to be performed without reconstructing the count. The servers use a garbled circuit (GC) to evaluate whether the secret-shared count satisfies the vulnerability predicate, revealing only a masked Boolean value of the vulnerability predicate. The mask prevents either server from learning the comparison result directly. We therefore arrange the output-sharing mechanism so that the result is delivered to the client, who is entitled to learn the vulnerability decision, while neither server obtains the result individually.

Our design admits two variants. The first avoids some computation by allowing the query's attribute-selection vector to be revealed to the servers, while the second hides this information at additional computational cost. We refer to these as $\Sigma_{VT^{-}}$ and $\Sigma_{VT}$, respectively. The detailed protocols and their trade-offs are given in \cref{sec:protocol}.

%The resulting construction therefore separates the query computation into three logically distinct components: private equality testing, secret-shared aggregation, and private threshold comparison. \Cref{sec:protocol} gives the complete protocol specification and describes the two construction variants and a worked example.

\section{Protocol Construction} \label{sec:protocol}

This section is divided into two parts. We first present $\Sigma_{\mathsf{VT}^{-}}$, a simpler variant that is linearly faster than the full construction but permits limited information leakage. We then present the full construction, $\Sigma_{\mathsf{VT}}$, which eliminates this leakage. Both variants comprise insertion and query protocols.

\subsection{$\Sigma_\mathsf{VT^-}$: Attribute-Selective Variant} \label{VT-}
This section presents two variants of our construction. The first, $\Sigma_{\mathsf{VT^-}}$, is an \emph{attribute-selective variant} that reveals which attributes are queried but hides their values, allowing
servers to avoid computation on unselected attributes and thus achieves a more efficient protocol. The second, $\Sigma_{\mathsf{VT}}$, is a \emph{query-private variant} that additionally hides the queried attributes to maximise user privacy. Both variants comprise insertion and query protocols.
%\paragraph*{$\Pi_\mathsf{VT^-}^\textit{Insert}$}
\paragraph{Insertion.}
To insert a record $v$, the client samples a random Boolean share $[v]_1$ and computes $[v]_2=[v]_1\oplus v$. The two shares are sent to the respective servers. The client is then not required to participate in subsequent queries. We arbitrarily reserve the vector of all 0s to mean undefined. When a user joins the system for the first time, they are asked for their personal data. The client then splits the vector into two shares $[v]^1$ and $[v]^2$ by first randomly choosing $[v]^1$ and then calculating $[v]^2 = [v]^1 \oplus v$, such that neither server can learn the user's data without both parts. The servers collect records from other clients in the same way. The formal protocol $\Pi_\mathsf{VT^-}^\textit{Insert}$ can be found in the \Cref{app:prot} as \Cref{app:insert-}.% The complete insertion protocol is given in \Cref{app:insert-}.

%At the start of the protocol, a client holds a \emph{record} which details information about themselves (e.g. age, gender etc.). The record is a string of bits $v$ formed by the concatenation of subvectors. Each subvector is of pre-defined length and corresponds to a certain attribute, while each possible value of the subvector corresponds to a certain attribute value. For example, if there are 10 age buckets, then the attribute of age must have a subvector of at least 4 bits, where each possible combination of bits corresponds to one age bucket (bar some invalid combinations). The mapping of the attributes and values and the exact structure of the record is defined at the implementation level. For example, for a system that uses attributes \{\emph{biological gender, age, sexuality, job}\} and a client who is a heterosexual woman of age 30-40 and unknown/undefined job has the following record:
%$$\overbrace{01}^{\text{gender}}|\overbrace{0100}^{\text{age}}|\overbrace{01}^{\text{sexuality}}|\overbrace{00000000}^{\text{job}}$$ 

%\paragraph*{$\Pi_\mathsf{VT^-}^\textit{Query}$}
\paragraph{Query.}
A query consists of an attribute-selection vector $q$ and a value vector $t$. In $\Sigma_{\mathsf{VT}}^{-}$, $q$ is sent to both servers in the clear, while $t$ is secret-shared as $[t]^1\oplus[t]^2=t$. Server~2 negates its share of $t$.

For each database record $v$, each server applies its query share as a bitmask to the corresponding record. The servers then perform distributed multiplication across the selected bits. The resulting Boolean shares $[z]^1,[z]^2$ satisfy
\[
z=[z]_1\oplus[z]_2 =
\begin{cases}
1 & \text{if the selected attributes of $v$ equal $t$,}\\
0 & \text{otherwise.}
\end{cases}
\]
Thus, one secret-shared equality bit is obtained for every database record. The equality results are converted to arithmetic shares using $\Pi_{\mathrm{B2A}}$ and locally summed. The resulting count remains secret-shared; reconstructing it would violate the privacy goal. For example, a server that knows an individual belongs to the database could issue a query fixing all known attributes and vary an unknown attribute across successive queries, thereby recovering the individual's value from exact population counts. We therefore evaluate only the predicate $\mathsf{count}<T$ using a garbled circuit, revealing the Boolean result but not the count.

Finally, the distributed multiplications are evaluated using the multiplication protocol described in \Cref{sec:prelims}. Multiplications can be arranged as a balanced tree, reducing the number of sequential communication rounds to $O(\log \ell)$ for an $\ell$-bit test vector. Concretely, $\Sigma_\mathsf{VT^-}$ consists of $\Pi_\mathsf{VT^-}^\textit{Query}$ and $\Pi_\mathsf{VT^-}^\textit{Insert}$, while $\Pi_\mathsf{VT^-}^\textit{Query}$ consists of the subroutines $\mathsf{\Pi}_{\textit{Data-prep}}$ (XOR data with $t$), $\Pi_\mathsf{Intersect^-}$ (equality testing), $\Pi_\mathsf{B2A}$ (share conversion) \cite{prio+}, and $\Pi_\mathsf{GC}$. The full protocol is detailed in \Cref{app:vt-}, and a walk-through use case is presented in \Cref{app:usecase}.

\paragraph{Leakage and efficiency.}
The primary limitation of $\Sigma_{VT^{-}}$ is that $q$ is revealed to the servers. This does not reveal the queried values, but may nevertheless leak information about the client's interests. In return, equality testing is performed only for selected attributes. Thus, if a query selects a fraction $\alpha$ of the attributes, the equality-testing work is approximately $\alpha$ of that required by the full construction. The precise benefit depends on the query workload. \Cref{sec:sigma} removes the leakage of $q$ at additional computational cost.

\subsection{$\Sigma_\mathsf{VT}$: Query-Private Variant} \label{sec:sigma}
Our protocol $\Sigma_\mathsf{VT}$ removes the aforementioned leakage of the query by secret-sharing $q$ and evaluating the equality test independently for each attribute.

To facilitate this, the construction requires a publicly reserved value for each attribute that cannot occur in a valid record. We arbitrarily use the all-ones vector $1^{\ell_i}$. This value acts as a guaranteed non-match when an attribute is excluded from a query. Consequently, its equality result can be used to neutralise the excluded attribute without revealing whether that attribute was selected. During insertion, the servers therefore reject any record containing the reserved value. This verification is performed using the same equality-testing primitive as the query protocol.

For a query $(q,t)$, the client secret-shares both the selection vector and the query values. For each attribute $i$, the servers receive shares corresponding to
\[
[q_i]\mathbin{|}[t_i]
\qquad\text{and}\qquad
\neg[q_i]\mathbin{|}\neg[t_i],
\]
respectively. The equality test is then performed separately on each attribute. The result is a Boolean share $[z_i]$ for each attribute.

For a selected attribute, $q_i=1$ and the equality test is evaluated on the client's supplied value $t_i$, so $z_i=1$ exactly when the database value matches the query. For an unselected attribute, the protocol replaces the query value with the publicly reserved all-ones value, which cannot occur in a valid database record and hence its equality test always returns $0$. The servers then incorporate the secret selection bit into the result, causing the final value for an unselected
attribute to be $1$. Thus, after this step, 
\[
z_i =
\begin{cases}
1 & \text{if attribute $i$ is not queried,}\\
1 & \text{if attribute $i$ is queried and matches,}\\
0 & \text{if attribute $i$ is queried and does not match.}
\end{cases}
\]
Multiplying the resulting shares across all attributes therefore produces $1$ exactly when the database record matches the query on every selected attribute. The remainder of the protocol is identical to $\Sigma_{VT^{-}}$: the resulting equality bits are converted to arithmetic shares, aggregated over the database, and compared privately with the threshold $T$.

The main cost of hiding $q$ is that equality testing must be performed for every attribute, including attributes that are not selected by the query. This removes the query-structure leakage of $\Sigma_{VT^{-}}$ at the cost of additional computation. The equality tests are also performed separately within each attribute, which may increase the number of sequential
distributed-AND operations. Nevertheless, using a balanced multiplication tree, the full construction requires at most one additional communication round compared with $\Sigma_{VT^{-}}$. Similarly to $\Sigma_\mathsf{VT}$, $\Sigma_\mathsf{VT}$ consists of $\Pi_\mathsf{VT}^\textit{Query}$ and $\Pi_\mathsf{VT}^\textit{Insert}$, while $\Pi_\mathsf{VT}^\textit{Query}$ consists of the subroutines $\mathsf{\Pi}_{\textit{Data-prep}}$ , $\Pi_\mathsf{Intersect}$, $\Pi_\mathsf{B2A}$ \cite{prio+}, and $\Pi_\mathsf{GC}$. The full protocol is detailed in \Cref{app:VT}

\subsection{Optimisations and Additional Features}
\paragraph*{Equality Testing.}
The equality testing (ET) protocol within our system can be considered as a black box and can easily be replaced by another one, so long as the result is returned as a Boolean share to each server. In particular, we consider the work of \cite{cout}, which showcases an equality testing protocol optimised for the offline phase. The protocol consists of two parts - an iterative size-reducing routine, and a routine for evaluating equality on small strings. We implement this alternative ET protocol, as well as a hybrid protocol that uses the size-reducing part of \cite{cout} and the normal ET protocol using distributed AND on the remaining small strings. All of these options are compared in the evaluation (see \Cref{eval}).

\paragraph*{Client Opt-out/Update.}
A practical addition to the system is the ability for a user to opt out of the system, such that their data be removed from the dataset. This can be realised using a digital signature scheme, where a client is issued a secret pseudonym. A client can present their pseudonym along with a valid signature when it wishes to opt out, and both servers can remove the record from the database. A client can effectively update their entry by removing it and then replacing it with a new one. 

%$$[\mathcal{R}_i]^1 \hspace{80pt} [\mathcal{R}_i]^2$$
%$$ [d]^1_{i1}, \overbrace{\ldots , [d]^1_{ik} | \text{dec}}^{[\mathcal{R}_i]^1}(x_i) \hspace{20pt} [d]^2_{i1}, \overbrace{\ldots , [d]^2_{ik} | \text{dec}}^{[\mathcal{R}_i]^2}(x_i)$$

\paragraph*{Parallel Preprocessing.}
The offline phase can be scaled horizontally by using additional non-colluding server pairs to generate Beaver triples in parallel and forward them to the two online servers. This increases preprocessing throughput without changing the online protocol, provided that the
additional servers satisfy the same non-collusion assumptions.
%\paragraph*{Beaver-triple Forwarding}\label{tripleforwaring}
%For large applications, there will inevitably be a bottleneck in the ability to generate enough Beaver triples (or the comparable preparation requirement for the 'Couteau' variant of ET). However, the generation of such is not explicitly limited to 2 machines, even in the face of 2PC protocols. That is, extra pairs of machines can generate triples and forward them to the main 2 servers without breaking the security requirement, so long as no additional machine colludes with its counterpart or the opposing main server. With such an infrastructure, the concern about the offline phase is more about computational expense rather than a bottleneck in performance (see \Cref{fig:4}). We make no direct efficiency claim - we simply have linearly more triples (in the number of extra pairs of servers) available in the same amount of time.

\subsection{Complexity Analysis}
We break down the complexity analysis based on the offline and online phases of the protocol. Here, the offline phase denotes all server-side computation that can be performed before the query is known, including Beaver-triple generation.
\paragraph*{Offline Phase.}
The cost of computation in the proposed system is dominated by the generation of Beaver triples. In the offline phase, privately computing the AND function over two random secrets (equivalent to multiplying over a Boolean field) has a cost of one OT per server. The number of multiplications required is $n * (l - 1)$, where $n$ is the number of clients and $l$ is the bitlength of the query. However, thanks to the work of Ishai \emph{et al}.~\cite{OTe}, it is known that a constant number of OTs can be `extended' to produce more OTs using cheap alternatives---namely invocations of a PRNG. Since many multiplications are needed (e.g. over 3 million for a 32-bit query over 100,000 clients), this optimisation has significant impact in this amortised setting. Ultimately, the complexity is dominated by the number of multiplications, $O(nl)$. Additionally, we saw multiplications optimised to reduce the number of round-trips, which is burdened with a tradeoff with space complexity. In particular, the offline phase requires $\lceil \text{log}_2(\sum_{i=1}^m2^{\lceil \text{log}_2l_i\rceil})\rceil$ round-trips and a space complexity dominated by copying the entire dataset. Note that it is quite possible to entertain this tradeoff in the case of having a very large dataset with small $l$ or round-trip delay, for example by computing over sections of the dataset in series, or by adjusting the order of multiplications.

\paragraph*{Online Phase.}

During the multiplication protocol, each server sends one bit to the other for each multiplication. This occurs after the servers mix their secrets with the Beaver triple, i.e. when they publish $[d]^i = [x]^i - [a]^i$ and $[e]^i = [y]^i - [b]^i$. Multiplication can be done mostly in parallel---namely, multiplications whose inputs do not require the result of a previous one. The total online communication is linear in the database size and query
length, i.e., $O(n\ell)$ bits, while the number of online rounds grows only logarithmically with the attribute bitlength. This separation is important for the intended setting: once records have been inserted and the preprocessing has been performed, queries can be executed without any interaction with the contributing clients. %Intuitively, the multiplications can be seen as a binary tree, where the root node is the server's share of the equality testing result.

%The number of rounds trips is the same as the offline phase and the total amount of data sent is $2n$log$_2l+n$ bits.
\begin{figure}
    \centering
    \includegraphics[width=1\linewidth]{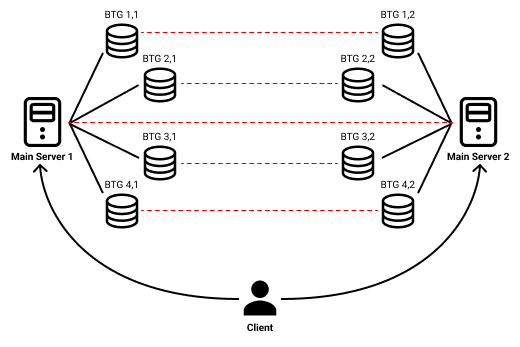}
    \caption{System architecture featuring additional optional BTG (Beaver Triple Generator) servers, arranged as above. If using 'Couteau' \cite{cout} for equality testing, the additional servers perform a similar routine requiring the same architecture}
    \label{fig:4}
\end{figure}

\section{Security Analysis} \label{sec:sec}
In this section we concretely analyse the security of $\Sigma_\mathsf{VT}$. We use a simulation-based proof strategy, in which we define what an ideal functionality means in our context, and define \emph{reactive} f-privacy, that captures dynamic databases, as is realistic in deployed settings.
\subsection{Security Model} \label{sec:model}
\paragraph{Threat Model.}
We adopt the threat model defined in Section~4.1. The adversary may corrupt at most one of the two servers and an arbitrary subset of clients. The servers are assumed to be semi-honest and non-colluding, while corrupted clients may deviate arbitrarily from the protocol. Communication between all parties is assumed to use secure channels.

Our privacy guarantee therefore protects the inputs of honest clients against a corrupted server, even when that server colludes with corrupted clients. Correctness and robustness against arbitrary malicious clients are guaranteed provided that neither server is corrupted. 
\paragraph{Reactive $f$-privacy.}

We now introduce an extended notion of $f$-privacy~\cite{prio,prio+} to account for the stateful nature of the system due to the dataset $\mathcal{D}$. Intuitively, $f$-privacy states that an adversary against a protocol computing the function $f$ is unable to learn any information beyond what is leaked by the evaluation of $f$. We extend this notion to \emph{reactive} $f$-privacy to additionally capture multiple evaluations over a changing data-set. More precisely, an adversary may observe the outputs of multiple, adaptively chosen queries over a changing dataset and therefore the privacy guarantee must hold across the entire sequence of insertions and queries, rather than only for a single query executed on a fixed dataset. This is particularly important in practical settings such as social media, where large numbers of users may issue repeated, adaptive queries over a continuously changing population.

\begin{definition}[Reactive $f$-privacy] \label{def:react-f-privacy}
A stateful protocol is reactively $f$-private if, for any adversary controlling at most one server and an arbitrary subset of clients, there exists a Probabilistic Polynomial Time (PPT) simulator $\mathcal{S}$ that, given the public parameters, the identities of corrupted participants, and the sequence of outputs permitted by the ideal functionality, produces a view computationally indistinguishable from the adversary's view in the real execution.
\end{definition}

\paragraph{Ideal Functionality and Leakage.}

We define the ideal functionality to capture exactly the information that may be revealed by the protocol. Intuitively, for a query $Q=(q,t)$ over dataset $\mathcal{D}$, the functionality computes the vulnerability predicate:
\[ f_T(\mathcal{D},Q) =
\begin{cases}
1 & \text{if fewer than $T$ records in $\mathcal{D}$ match $Q$},\\
0 & \text{otherwise}.
\end{cases}
\]
The resulting bit is returned only to the querying client; neither server receives any query-dependent output.

For insertion, the functionality accepts a record provided that no attribute contains the reserved all-ones, $1^\ell$, value used by $\Sigma_{\mathsf{VT}}$. The validity result is revealed to the servers, since they must determine whether the record can be inserted. For an honest client, this result is always valid and therefore reveals no information about the record. 

Thus, apart from public parameters and the identities of corrupted parties, the ideal functionality reveals only the sequence of threshold results returned to clients and the validity of submitted records. In particular, it reveals neither the contents of the database nor the selection vector $q$, query values $t$, or intermediate equality-test results.

The protocol ideal functionality $\mathfrak{I}_{\mathsf{VT}}$ comprises of two functions representing the actions of a user: $\mathfrak{I}_{\text{Insert}}$ and $\mathfrak{I}_{\text{Query}}$. We start by describing the ideal functionality of the Insert algorithm.  Formally defined in \Cref{def:ideal_insert} functionality takes as input a client vector $v$ and returns $c=1$ if v contains the all-ones vector, and 0 otherwise. The outcome of this functionality is given to the servers, and not the client. If $c=0$, then the vector $v$ is added to the database.

Then, for a client query,  we define the ideal functionality $\mathfrak{I}_{\text{Query}}$ in \Cref{def:ideal_query}. This is a stateful protocol that receives the negation of the client's query $\neg q|\neg t$ and the stateful dataset $\mathcal{D}$ held jointly by the servers. It computes $b = 1$ if and only if the count of the query is less than the public threshold $\mathcal{T}$, and returns $b$ to the client only. The servers receive no output. 

%We now formally define the ideal functionality of $\Pi_\mathsf{VT}^\textit{Query}$, $\Pi_\mathsf{VT}^\textit{Insert}$,
%Formally, we define the ideal functionality of the insert and query functions below. 
\begin{definition} \label{def:ideal_insert}
    ($\mathfrak{I}_{\text{Insert}}$). The ideal functionality
    $\mathfrak{I}_{\text{Insert}}$ behaves as follows:
    \begin{itemize}
        \item $\mathfrak{I}_{\text{Insert}}$ receives the client's
        submitted vector $v$ (via its Boolean shares $[v]^i$ from
        server $i$, $i \in \{1,2\}$).
        \item $\mathfrak{I}_{\text{Insert}}$ computes the bit
        $c = 1$ if $v$ contains an all-one subvector (i.e. $v$ is
        malformed), and $c = 0$ otherwise.
        \item $\mathfrak{I}_{\text{Insert}}$ returns $c$ to both
        servers. The client receives no output.
        \item If $c = 0$, $\mathfrak{I}_{\text{Insert}}$ updates the
        stateful dataset $\mathcal{D} \leftarrow \mathcal{D} \cup
        \{v\}$; if $c = 1$, $\mathcal{D}$ is left unchanged.
    \end{itemize}
\end{definition}
\begin{remark}
Note that $c$ never leaks information about an honest client's private data: for an honest client, $c = 0$ deterministically, independent of $v$, since honest clients never submit malformed vectors. For a
malicious client, $c$ is already known to the adversary. In both cases, no information about $\mathcal{D}$ or about any \emph{other} client's data is revealed by $c$.
\end{remark}
\begin{definition} \label{def:ideal_query}
    ($\mathfrak{I}_{\text{Query}}$). The ideal functionality $\mathfrak{I}_{\text{Query}}$ behaves as follows:
    \begin{itemize}
        \item $\mathfrak{I}_{\text{Query}}$ receives the negation of the client's query $\neg q|\neg t$ and the stateful dataset $\mathcal{D}$ held jointly by the servers.
        \item $\mathfrak{I}_{\text{Query}}$ counts how many entries in $\mathcal{D}$ are equal to the client's query (formed with $q|t$), and computes $b = 1$ if and only if the count of the query is less than the public threshold $\mathcal{T}$.
        \item $\mathfrak{I}_{\text{Query}}$ returns $b$ to the client only.
    \end{itemize}
\end{definition}

The ideal functionality exposes only the information that is necessarily available to the corresponding parties. The servers learn the public parameters and, for each insertion, the validity bit $c$; they receive no query-dependent output. A querying client learns only the threshold bit $b$. A malicious client additionally knows its own submitted record or query and may adapt its future actions based on the
outputs it receives. No party learns the honest clients' records, the query values or selection vector of an honest querying client, or any intermediate equality-test or aggregation values. The sequence of insertion-validity bits and query-output bits constitutes the complete non-public leakage permitted by $\mathfrak{I}_{\mathsf{VT}}$.

\begin{remark}
    \Cref{def:react-f-privacy} characterizes the leakage of each query as exactly the threshold bit for the dataset state at that time. When the dataset changes between two queries of the same $(q,t)$ pair, an adversary can combine the two outputs to learn whether an intervening change moved the count across the threshold---information not obtainable from either query alone. This is an instance of the differencing/reconstruction phenomenon well known in the aggregate statistics literature (e.g. \cite{differencing}), and is an inherent consequence of offering exact, deterministic query answers over a mutable dataset; it is not specific to $\Sigma_\mathsf{VT}$ and would apply to Prio-like schemes under a reactive dataset model. In fact, our construction leaks strictly less information by only revealing the result of the threshold test, rather than the exact sum. Mitigating this consequence further (e.g. via differentially-private noise addition, rate-limiting, or restricting repeat queries) is out of scope for this work, which targets exactness and cryptographic (not statistical/output) privacy.
\end{remark}
\subsection{Security Proof} \label{sec-proof}

We prove security in two stages. First, we show that a single query or insertion can be simulated using only the outputs specified by the ideal functionalities. We then extend this result to an arbitrary sequence of adaptive insertions and queries using a standard hybrid argument.

\begin{lemma}[\textbf{Single-round privacy}]
\label{lem:single-round}
For any single query invocation, $\Pi_\mathsf{VT}^\textit{Query}$ is computationally indistinguishable from an execution of ideal functionality $\mathfrak{I}_{\text{Query}}$ against the adversary structure of \Cref{sec:model}.
\end{lemma}
\begin{proof}

We now build an efficient simulator $\mathcal{S}$ for $\Pi_\mathsf{VT}^\textit{Query}$ and show it is indistinguishable from the real protocol. The protocol begins with a local secret XOR and secret multiplication. Since the XOR only incorporates local secrets, we argue that each share that any corrupted party sees is uniformly random and independent of the underlying secret. As such, the simulator simply samples shares uniformly at random. Then, the Beaver multiplication protocol securely realises the ideal multiplication functionality \cite{beaver}, so this step may be treated as a call to $\mathfrak{I}_{\text{Mult}}$ (defined in \Cref{app:ideals}). As such, the simulator also samples uniformly at random for these outputs. 

Similarly, the Boolean-to-arithmetic conversion protocol \cite{prio+} securely realises $\mathfrak{I}_{\text{B2A}}$, so the simulator can once again sample uniformly at random. For step 4, the simulator uses the ideal functionality ($\mathfrak{I}_{\text{GC}}$). Such a construction is proven to be secure by Lindell and Pinkas~\cite{gc}. In the real protocol, the adversary learns 1) $b \oplus r$ if controlling server 1 and 2) $r$ if controlling server 2. For 1), the simulator outputs $\hat{r}$ by calculating $b \oplus (b \oplus r)$, and for 2), the simulator already knows $r$. Thus, the output of the simulator preserves the mathematical relation.

\begin{figure*}
    \centering
    \includegraphics[width=0.8\linewidth]{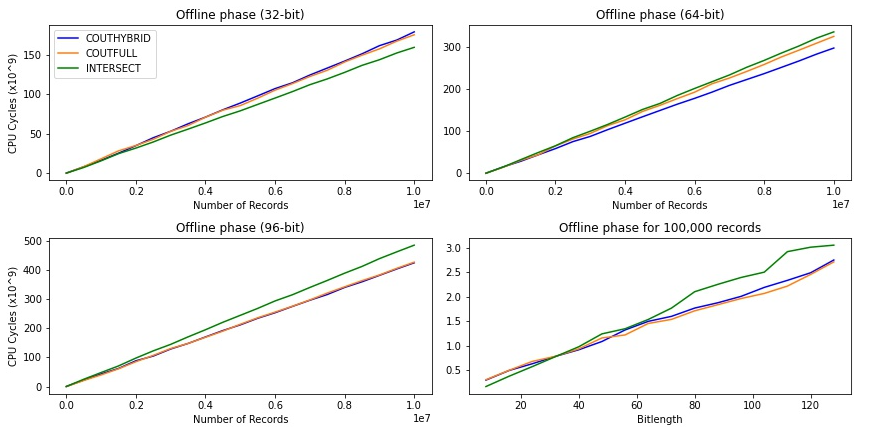}
    \caption{Benchmarks for offline phase for INTERSECT, COUTHYBRID, and COUTFULL}
    \label{fig:plots1}
\end{figure*}

To summarise, by the standard composition \Cref{lem:composition}, and since Beaver's protocol securely realises $\mathfrak{I}_{\text{Mult}}$ \cite{beaver}, the share conversion protocol of Addanki \emph{et al.}~\cite{prio+} securely realises
$\mathfrak{I}_{\text{B2A}}$, and semi-honest Yao's protocol securely realises $\mathfrak{I}_{\text{GC}}$ \cite{gc}, the $\Pi_\mathsf{VT}^\textit{Query}$ realises the ideal functionality and is indistinguishable from the real protocol.
\end{proof}

\begin{corollary}[\textbf{Insert privacy}]
\label{cor:insert}
For any single insert invocation, $\Pi_\mathsf{VT}^\textit{Insert}$ is computationally indistinguishable from an execution of ideal functionality $\mathfrak{I}_{\text{Insert}}$ against the adversary structure of \Cref{sec:model}.
\end{corollary}
\begin{proof}
The verification subroutine of $\Pi_\mathsf{VT}^\textit{Insert}$ invokes exactly the same machinery as steps 1-2 (secret sharing and Beaver multiplication) analysed in \Cref{lem:single-round}; by the
identical argument, each share seen by a corrupted party is uniformly random and independent of the underlying secret, so the simulator samples these uniformly, treating Step~2 as a call to
$\mathfrak{I}_{\text{Mult}}$ as before.

Unlike the query flow, the resulting comparison bit here is revealed in the clear directly between the two servers, rather than routed through a garbled circuit visible only to the client. Since one party's share of an additively/XOR-shared value is uniformly random and independent of the secret, the simulator samples a uniformly random share for the honest party's message and sets the corrupted party's share to be consistent with the known value of $c$ (obtained from $\mathfrak{I}_{\text{Insert}}$). By \Cref{lem:composition}, composing this with the $\mathfrak{I}_{\text{Mult}}$-hybrid argument for Steps 1-2 gives real-model security.
\end{proof}

\begin{theorem}[Reactive $f_T$-privacy]
$\Pi_{\mathsf{VT}}$ satisfies reactive $f_T$-privacy against the adversary defined in \Cref{sec:model}, i.e. is computationally indistinguishable from the ideal functionality $\mathfrak{I}_{\mathsf{VT}}$.
\end{theorem}
\begin{proof}

Let $N$ be the number of query/insert invocations in the run. Define hybrids $G_0, \dots, G_N$, where $G_r$ simulates the first $r$ invocations and executes invocations $r+1, \dots, N$ as in the real protocol. The simulator maintains the same dataset state as the ideal functionality and processes each invocation in order. For each query it uses only the threshold bit returned by $\mathfrak{I}_{\mathsf{Query}}$, while for each insertion it uses only the validity bit returned by $\mathfrak{I}_{\mathsf{Insert}}$.

Since each round's randomness (shares, Beaver/B2A pre-computation, GC garbling) is freshly and independently sampled, \Cref{lem:single-round} gives $\mathsf{Adv}(G_{r-1}, G_r) \le \varepsilon_r$ for negligible $\varepsilon_r$, for each $r$ where $\mathsf{Adv}$ denotes the advantage in distinguishing its inputs. By the triangle inequality, $\mathsf{Adv}(G_0, G_N) \le \sum_{r=1}^N \varepsilon_r$, which is negligible since $N$ is polynomial in the security parameter. Hence $G_0$ (fully simulated) and $G_N$ (fully real) are indistinguishable, as required.
\end{proof}
\subsection{Correctness} \label{correctness}
We establish the correctness based on the cryptographic building blocks.
\begin{theorem}[Correctness]
For a dataset $D$ and query
$(q,t)$, $\Pi^{\mathsf{Query}}_{\mathsf{VT}}$ outputs
\[
b = f_T(D,q,t)
\]
to the querying client except with negligible probability.
\end{theorem}
\begin{proof}[Proof]

We prove correctness for honest servers when the dataset contains only valid records, i.e. are accepted by $\Pi_\mathsf{VT}^\textit{Insert}$. We begin with the observation that an AND function on multiple inputs evaluates to 1 if and only if all inputs are 1, by definition. By the same logic, a bitstring evaluates to 1 when all its bits are multiplied together, if and only if all bits are set to 1. We also observe the commutative and associative property of $\oplus$, such that $(x \oplus y) \oplus (z \oplus w) \equiv (x \oplus z) \oplus (y \oplus w)$ for vectors (or bitstrings) of equal length. Since Boolean shares are built from XORs ($x = [x]^1 \oplus [x]^2$), it is easy to verify that, for a secret-shared record $\mathcal{R} = [\mathcal{R}]^1 \oplus [\mathcal{R}]^2$ and secret-shared query (test vector) $t = [t]^1 \oplus [t]^2$, one can test for equality by taking the negation of the test vector $\neg t = [t]^1 \oplus \neg [t]^2$ and XORing the corresponding shares, $([t]^1 \oplus [\mathcal{R}]^1) \oplus (\neg[t]^2 \oplus [\mathcal{R}]^2) = ([t]^1 \oplus \neg [t]^2) \oplus ([\mathcal{R}]^1 \oplus [\mathcal{R}]^2) =  \neg t \oplus \mathcal{R} = $ the vector of all 1s if and only if $t = \mathcal{R}$. Therefore, it suffices to perform the equality test by performing distributed AND on the masked records, i.e. $[\mathcal{R}]^1 \oplus [t]^1$ for server 1 and $[\mathcal{R}]^2 \oplus \neg[t]^2$ for server 2. Proof of correctness for distributed multiplication using Beaver Triples is found in \Cref{triples}.

At the conclusion of the distributed multiplication protocol, each server has one one-bit share of the equality predicate $[z]^{\{1,2\}}$ per record. At this point, the servers would like to aggregate homomorphically, that is, $([z]_1^1 +_l \ldots +_l [z]_n^1) +_l ([z]_1^2 +_l\ldots +_l [z]_n^2) \leftrightarrow ([z]_1^1 +_2 [z]_1^2) +_l \ldots +_l ([z]_n^1 +_2 [z]_n^2)$ where $+_l$ indicates addition in the field of $\mathbb{F}_l$. %In order to do so, the elements of $z$ must be converted to $l$, and the field $\mathbb{F}_l$ must not be overflowed in order to preserve correctness. 
The equality bits must be converted from $z$ to an arithmetic field \(\mathbb F_p\) with \(p>l\), ensuring that the aggregate count does not wrap modulo \(p\). This is due to a classic observation popularised by \cite{prio}. Performing this conversion is a similar process to using Beaver Triples, and the proof of correctness can be found in \Cref{B2A}. Finally, the system concludes by having the servers privately test whether their homomorphic aggregate shares will exceed the threshold when combined. This can be done straightforwardly with garbled circuits, a standard technique whose proof of correctness is omitted. Hence the output is equal to $f_T(D,Q)$ (recall $Q=(q,t)$), except with negligible probability due to the underlying cryptographic primitives.

The addition of $q$ is purely semantic. In the view of the client, it is used to indicate whether a section of the query should be considered or not, which is equivalent to letting it always pass the ET for that section. By taking that result and XORing it by $\neg q$, mathematically speaking, this simply computes the truth value of whether the corresponding section of $t_*$ was \emph{not} in the record. Paired with letting the client set $t_*$ to a subvector that is \emph{never} in the record, this achieves the semantics required of the client above.

%For each record $v\in D$, the equality-testing protocol outputs a secret-shared bit indicating whether each queried attribute matches the corresponding attribute of $v$. For attributes not selected by $q$, the query masking sets the corresponding comparison value to the reserved all-ones value. Since this value is excluded from valid records,
%unselected attributes contribute $1$ to the equality test. Consequently, the product of the per-attribute equality results is $1$ exactly when $v$ satisfies all attributes selected by $q$.
%
%The protocol then sums these secret-shared record-level matching bits, obtaining a secret share of the number of records satisfying $(q,t)$. Finally, the threshold comparison evaluates $f_T$ on this shared count and reveals only the resulting bit to the querying client.
\end{proof}
\section{Implementation and Evaluation} \label{eval}

\begin{figure*}
    \centering
    \includegraphics[width=0.8\linewidth]{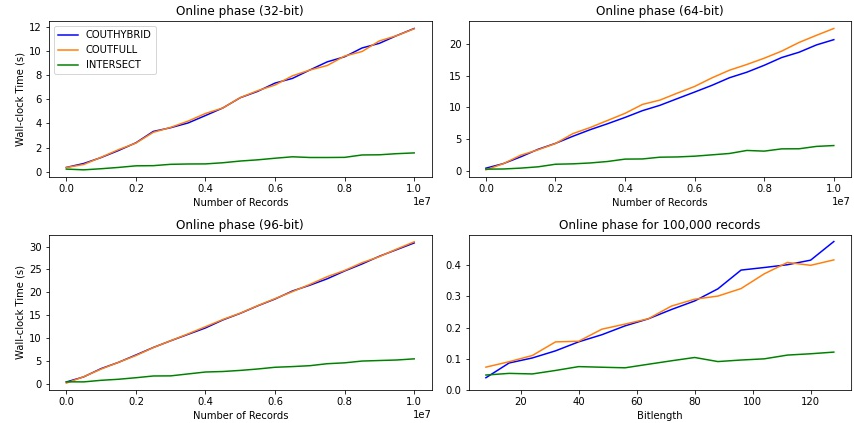}
    \caption{Benchmarks for online phase for INTERSECT, COUTHYBRID, and COUTFULL}
    \label{fig:plots2}
\end{figure*}

The goal of the following section is to present and discuss the viability of the real system we have built. We aim to compare the main scheme $\Sigma_\mathsf{VT}$ based on `intersect', the distributed AND function, with the alternative system that uses the equality testing protocol `Couteau' \cite{cout} as well as the hybrid method (using the size-reducing routine within \cite{cout} followed by `intersect' on the reduced strings) in both the offline and online phase. For the evaluation we call these INTERSECT, COUTFULL and COUTHYBRID respectively.

\noindent\textbf{Implementation}
We implement the system in C and C++, using the Sodium library in C for random number generation, libOTe in C++ for OT extensions \cite{libOTe}, and the emp-sh2pc library for the garbled circuit \cite{emp}. The implementation is available at \url{https://anonymous.4open.science/r/privateintersect-F711/}. 

The implementation includes 3 programs: server 1, server 2, and the client. Communication is properly constructed using sockets such that all programs can be run from any machine and target any IP and port, and any number of clients can connect. We also implement the other variants of the protocol: COUTFULL, COUTHYBRID, and INTERSECTMINUS which is the alternative scheme presented in \Cref{VT-} (for INTERSECT only). The implementation is optimised for space and can thus run queries of up to 10 million records and 128 bits of query length with just 8GB of RAM. We leverage space optimisations such as efficient data storage, array reuse, and performing the data-prep roles simultaneously inside $\Pi_\mathsf{Intersect}$.

We present benchmarks for 32-bit, 64-bit, and 96-bit queries for the 3 variants of ET in the offline phase (\Cref{fig:plots1}) and online phase (\Cref{fig:plots2}) --- showing how efficiency scales with the number of records that are computed over in the dataset. We include an additional plot that shows how efficiency scales with bitlength for a fixed number of records. 

\noindent\textbf{Hardware} We tested our system on Linux Debian 12 using 2 OpenNebula VMs that are on physically separate machines. Both machines possess an Intel(R) Xeon(R) Silver 4316 CPU@2.3GHz, while the VMs ran at 2.2GHz. Machine 1 ran server 1 and the benchmarks and Machine 2 ran server 2 and a client. As such, the client did not benefit from being locally connected to server 2. Although the setup exhibited low latency $<1$ms, we did not leverage any special data-centre optimisations. CPU cycles were measured using the hardware performance counter 
\begin{verbatim}
    perf stat -e cycles -p <IDs>
\end{verbatim}
where IDs is the list of threads that do computation (i.e. excludes socket listener thread).

\noindent\textbf{How to Read the Data}
Online cost is measured in wall-clock time. That is, the realistic time for the measured code to complete. We use this metric because much of the cost comes from round trip latency. One can estimate the response time for a high latency setup by multiplying the number of round trips for a certain query (see \Cref{rtt}) by the given latency and adding it to the corresponding data point. 

Offline cost is measured in CPU cycles per server. We use this metric due to the facts that 1) computational cost dominates and 2) pre-computation is infinitely parallelisable up to the granularity of 1 Beaver triple (or equivalent for Couteau ET) --- we can achieve practically any completion time for a given query using sufficient numbers of Beaver-Triple servers to pre-generate enough Beaver triples for the observed query volume. To estimate the CPU time for a data point, we can simply divide by the clock speed of the CPU. For the sake of comparison, we include \Cref{fig:plots3} in \Cref{app:furtherbenchmarks} which shows plots for offline phase in wall-clock time and online phase in CPU cycles.

\noindent\textbf{Client Cost}
Clients suffer little cost, as they are required only to generate shares of their data $[v]^i$ for $\Pi_\mathsf{VT}^\textit{Insert}$ and $[q]^i|[t]^i$ for $\Pi_\mathsf{VT}^\textit{Query}$, for the 2 servers $i \in \{1,2\}$. The client must compute a single 1-bit XOR at the end of $\Pi_\mathsf{VT}^\textit{Query}$.

%\paragraph*{Server Costs.}
\noindent\textbf{Online Server Costs}\ Figure~\ref{fig:plots2} shows online query latency as the database size up to $10^7$ records. At $10^6$ records for $\ell = 32$, the complete online query requires $\approx 2s$, $\approx 2s$ and $\approx 0.1s$ of computation for COUTHYBRID, COUTFULL and INTERSECT respectively. Increasing $n$ from $1{,}000{,}000$ to $10{,}000{,}000$ increases latency time by $\approx 10\times$, consistent with the $O(n\ell)$ complexity of the construction. Our results show our INTERSECT protocol dominate in the online phase. We observe that the performance gap of INTERSECT initially is more performent when the bitlength is $32$, but gives way to COUTHYBRID and COUTFULL at approximtely $\ell = \approx 60$, which is consistent with the fact that \cite{cout} is a recursive, logarithmic shrinking function. 

\noindent\textbf{Offline Server Costs} For $n=1{,}000{,}000$ records and $\ell=32$, preprocessing requires $n(\ell-1)=31{,}000{,}000$ Boolean multiplications. In our implementation, this corresponds to $\approx 2 \times 10^{10}$ cycles for all COUTHYBRID, COUTFULL and INTERSECT. This increases to $16\times10^{10}$ cycles when $n=10{,}000{,}000$, again consistent with the $O(n\ell)$ complexity of the construction. As with the online phase, we observe that the performance gap of INTERSECT decreases marginally as the bitlength increases. We conclude that INTERSECT is overall superior for realistic bitlengths used in our targetted context, but note that the COUTFULL approach could be desirable in the case of very large bitlengths.

\noindent\textbf{$\Pi_\mathsf{VT}^\textit{Insert}$ Cost}
Results for $\Pi_\mathsf{VT}^\textit{Insert}$ can be found by taking the lowest data point on its line, since all the verification is doing is the main protocol on only 1 client.

\subsection{System Goals Assessment}
In this section we revisit the system goals from \Cref{reqs} and assess whether they have been satisfied. 

\noindent\textbf{Robustness} 
Recall that the system is robust if an adversary cannot corrupt the system any more than misreporting their information. An adversary often achieves this goal in private systems that use an arithmetic secret-sharing scheme, since the servers cannot trivially verify the validity of the submission without breaking the privacy of the user. Such systems must then use rather complex techniques to maintain robustness. On the contrary, Prio+~\cite{prio+} and indeed this work enjoy the benefit of using a Boolean secret-sharing scheme. That is, each attribute has a pre-defined length. The servers can then reject any submissions that deviate from the expected length. Note that this means that the range of values is only enforcible upon powers of 2. Since the submission is the only point the client can influence the server, the robustness argument is complete. 

%\paragraph*{Security.}
%We proved security in \Cref{sec}, so we take this opportunity to discuss security-related design choices. 
%This paper is heavily inspired by Prio, and we agree on many of their design choices and reasoning. In particular, protecting against malicious clients appears to be a minimum for a usable system. Upon deciding on using the approach of $\Pi_\mathsf{VT}^\textit{Query}$, we knew the routine would be relatively expensive since all the data in the dataset must be computed over for every query (without caching). As such, we have not considered extending the security notion to fully malicious non-colluding servers, as this would considerably increase overhead. That being said, such a construction may be appealing as future work if other advancements are found which speed up the current scheme. A similar argument applies to guaranteeing correctness in the case of faulty servers, which would concretely hinder performance. 

\noindent\textbf{Scalability}
We now assess scalability by considering the limit of what could be considered a reasonable cost for offline and online phase. Let us consider an example scenario of a medium-sized dataset of 1,000,000 records and a moderate 64-bit query. The example setup has 4 pairs of pre-computation servers, working at 4.5GHz. Then, our results show an offline phase of 1.8 seconds and an online phase of 0.38 seconds. If the pairs of servers have 50ms latency between them, the online phase would still only be $0.38 + $ (log$_2(64) + 1) \cdot 0.05 = 0.73$ seconds.

\noindent\textbf{Offline}
Despite being a very restrictive constraint, the offline property comes simply from the fact that data is fetched from a database stored by the servers rather than the users, which clearly does not require the users to be online. 

\subsection{Adequacy of Bitlength}
In this section, we analyse the bitlength required for realistic queries. The appropriate bitlength depends on the number and granularity of attributes: values should be sufficiently fine-grained to capture meaningful differences between individuals, while avoiding unnecessary domain sizes. For example, age may be represented in 10-year buckets
rather than individual years when this better reflects the information available to an adversary. Excessive generalisation, however, reduces the usefulness of the resulting uniqueness estimates. Thus, bitlength is ultimately a semantic design choice determined by the intended linkage-attack model.

There is no clear empirical basis for determining a universally appropriate bitlength; existing work on characterising individuals typically uses more sophisticated, AI-based similarity measures rather than discrete attributes. We therefore consider two representative case studies.

In the work of \cite{jangra}, Jangra et al. extract 19 PII-revealing attributes to study user oversharing on Reddit. Each attribute varies greatly in the number of values, ranging from just 2 for 'gender' and thousands for `location'. We estimate (through communication with the authors) an average of roughly $2^7$ values per attribute for a total of 133 bits. \cite{reddust} on the other hand characterise Reddit users with only 5 attributes, using binary predicates (1 bit) for gender and family status, 90 values of age (7 bits), 832 values for profession (10 bits) and 336 values for hobby (9 bits), for a total of 28 bits. The authors also discuss levels of generalisation for profession and hobbies and perform analysis on the correlation between attributes by generalising profession to only 30 values (5 bits), suggesting that the total bits required could be reasonably reduced to just 19 bits (even fewer if generalising age into buckets).

Ultimately, the choice of maximum query bitlength is a balance between usefulness and efficiency. We believe that the case studies combined with the performance analysis suggest that a  bitlength of 16, 32, 64 or 128 are all strong choices, and the optimal of these will depend on the number of records in the dataset as well as more nuanced implementation requirements.

\section{Discussion and Future Work}

%The current work as well as the state of the field of Privacy Preserving Measurements (PPM) showcase quite handily the struggle to create technologies that will realistically be used in industry. Privacy creates a huge overhead, especially when flexibility is required. Contributors to the field must work exhaustively to convince potential implementers that the privacy is worth the overhead.

%Despite many small advances in the field, it would seem likely that big tech companies will ignore all but the most efficient (and least flexible) privacy enhancing technologies without either a major breakthrough in the field or considerably increased pressure for data privacy. We do note, however, that the technology detailed in this work has far less problematic overheads for small-scale applications, and thus it is perhaps smaller companies for which we should be aiming, so to speak. 
Despite continued incremental advances, the overhead of privacy-enhancing technologies may limit their adoption in large-scale platforms, where efficiency and scalability are critical requirements. In the absence of substantial advances or stronger incentives for privacy-preserving data practices, highly efficient solutions may therefore be more likely to see widespread deployment. 

In our context, the system goal of `offline users' proved challenging for this work, as it excludes many successful lines of work, such as Verifiable Distributed Aggregation Functions (VDAFs) \cite{VDAF} and sketching~\cite{sketch1,sketch2,sketch3} protocols. In particular, we are not aware of any works in the field of PPM which explore the property of offline clients.

Nonetheless, our results suggest show that the overhead of the proposed system is considerably less restrictive for smaller-scale applications, where the flexibility of supporting arbitrary post-hoc queries may offer a more attractive trade-off. Our future work will focus on further improving performance; for instance,  leveraging intermediate values  for new computations without needing to begin from scratch. An important limitation of this work is the need to compute from scratch for every new query when using the base scheme. This could be mitigated with a caching system and would provide meaningful efficiency improvements.

\subsection{Conclusion}

We presented $\Sigma_{\mathsf{VT}}$, a privacy-preserving vulnerability testing system that allows a client to privately determine whether their data is sufficiently rare to pose a data-linkage risk. The key challenge is supporting arbitrary, post-hoc queries while allowing clients to be offline after contributing their data. We address this using secret-shared database records and a two-server computation that reveals only the threshold predicate, providing reactive $f_T$-privacy against a semi-honest server and malicious clients.

We introduced two constructions that trade query-structure privacy for efficiency, and showed how Beaver-triple preprocessing, parallelised multiplication, and alternative equality-testing techniques reduce the cost of the full construction. We introduced a dynamic notion of security, reactive $f$-privacy, and proved that our construction meets based on the ideal functionality of its building blocks. Our implementation demonstrates that the approach scales to datasets of up to 1 million records while retaining practical online query latency, with further parallelism available for the offline phase.

More broadly, this work demonstrates that  private vulnerability testing can be performed without requiring users to remain online or disclose their query or exact population count. We view the resulting system as a foundation for more efficient privacy-preserving testing of data-linkage risks, particularly in settings where the flexibility of post-hoc queries and the ability to support offline users are important.
%In this paper we introduce a private vulnerability testing tool that can aid a vulnerable user by informing them that they fall into a small minority and are thus vulnerable to data linkage attacks, which in turns means they should be more cautious about oversharing. Within this work, we provide the first formal solution to this kind of problem, laying groundwork for future optimisations. We propose our own optimisations and additional features and develop an implementation of the system described herein. We prove its security and viability for real-world deployment.

\bibliographystyle{ACM-Reference-Format}
\bibliography{alex1bib}

\appendix

\section{Full Protocol Description} \label{app:prot}

\subsection{$\Sigma_\mathsf{VT^-}$} \label{app:vt-}
$\Sigma_\mathsf{VT^-}$ consists of 5 parts. $\mathsf{\Pi}_{\textit{Data-prep}}$, $\Pi_\mathsf{Intersect^-}$, $\Pi_\mathsf{B2A}$ \cite{prio+}, $\Pi_\mathsf{GC}$, which together form $\Pi_\mathsf{VT^-}^\textit{Query}$, then additionally $\Pi_\mathsf{VT}^\textit{Insert}$. We present the pseudocode for each sequentially, excluding $\Pi_\mathsf{B2A}$.

\begin{algorithm}
	\caption{$\mathsf{\Pi}_{\textit{Data-prep}}$} 
	\begin{algorithmic}[1]
    \State Input: dataset structure $\sigma$, dataset share per server $[\mathfrak{D}]^{\{1,2\}}$, client query share per server $q|[t]^{\{1,2\}}$, algorithm choice $\mathfrak{A}$, bank of Beaver triples $[T]^{\{1,2\}}$
    \State Output: prepped data $[c]^{\{1,2\}}$
    \For{all $\mathcal{R}_i \in \mathfrak{D}$}
        \State $a \leftarrow 1$
        \For{$j \in \{1\ldots k\}$}
            \If{$q[j] = $ True}
                \State $[c]_{i,a}^x \leftarrow [d]_{i,j}^x \oplus [t]_a^x$ 
                \If{$\mathfrak{A} = $'intersect'}
                    \State $[c]_{i,a}^2 \leftarrow \neg[c]_{i,a}^2$
                \EndIf
                \State $a \leftarrow a + 1$
            \EndIf
        \EndFor
    \EndFor
    \State \Return $[c]^{\{1,2\}}$
	\end{algorithmic} 
\end{algorithm}

\begin{algorithm}
    \caption{$\Pi_\mathsf{Intersect^-}$}
    \begin{algorithmic}[1]
        \State Input: data shares per server $[c]^{\{1,2\}}$, query bitlength $|t|$, choice of algorithm $\mathfrak{A}$, bank of Beaver triples $[T]^{\{1,2\}}$
        \State Output: share of equality test results $[z]^{\{1,2\}}_i$ per server per record
        \If{$\mathfrak{A} = $"Couteau"}
            \State Employ Private ET scheme detailed in [7]
        \ElsIf{$\mathfrak{A} = $"Hybrid"}
            \State Employ Private ET scheme detailed in [7] but replace `product-sharing' function with $\mathfrak{A} = $"intersect" on the size-reduced strings
        \Else
            \State $R \leftarrow \lceil\text{log}_2(|t|)\rceil$
            \State $l \leftarrow |t|$
            \State $j_1 \leftarrow 1$
            \For{$r \in \{1\ldots R\}$}
                \State $P \leftarrow \lfloor l / 2 \rfloor$
                \State $l \leftarrow l - P$
                \State $s \leftarrow 2^r$
                \State $j_2 \leftarrow j_1$
                \For{$p \in \{1\ldots P\}$}
                    \State $i_1 \leftarrow 2ps$
                    \State $i_2 \leftarrow s(2p + 1)$
                    \For{all $\mathcal{R}_i \in \mathfrak{D}$}
                        \State $[D]^1_{i,p} \leftarrow [c]^1_{i,i_1} \oplus [T]^1_{j_1,1}$
                        \State $[E]^1_{i,p} \leftarrow [c]^1_{i,i_2} \oplus [T]^1_{j_1,2}$
                        \State $[D]^2_{i,p} \leftarrow [c]^2_{i,i_1} \oplus [T]^2_{j_1,1}$
                        \State $[E]^2_{i,p} \leftarrow [c]^2_{i,i_2} \oplus [T]^2_{j_1,2}$
                        \State $j_1 \leftarrow j_1 +1$
                    \EndFor
                \EndFor
                \State $j_1 \leftarrow j_2$
                \State Servers publish D and E shares
                \For{$p \in \{1\ldots P\}$}
                    \State $i_1 \leftarrow 2ps$
                    \State $i_2 \leftarrow s(2p + 1)$
                    \For{all $\mathcal{R}_i \in \mathfrak{D}$}
                        \State $D_{i,p} \leftarrow [D]^1_{i,p} \oplus [D]^2_{i,p}$
                        \State $E_{i,p} \leftarrow [E]^1_{i,p} \oplus [E]^2_{i,p}$
                        \State $[c]^1_{i,i_1} \leftarrow D_{i,p}\cdot [T]^1_{j,2} \oplus E_{i,p}\cdot [T]^1_{j,1} \oplus [T]^1_{j, 3}$
                        \State $[c]^2_{i,i_1} \leftarrow D_{i,p}\cdot E_{i,p} \oplus D_{i,p}\cdot [T]^2_{j,2} \oplus E_{i,p}\cdot [T]^2_{j,1} \oplus [T]^2_{j,3}$
                        \State $j_1 \leftarrow j_1 + 1$
                    \EndFor
                \EndFor
            \EndFor
            \For{all $\mathcal{R}_i \in \mathfrak{D}$}
                \State $[z]^{\{1,2\}}_i \leftarrow [c]^{\{1,2\}}_{i,1}$
            \EndFor
            \State \Return{$[z]^{\{1,2\}}$}
        \EndIf
    \end{algorithmic}
\end{algorithm}

\begin{algorithm} \label{app:gc}\caption{$\Pi_\mathsf{GC}$} \begin{algorithmic}[1] \State Input: Arithmetic share $[a]^{\{1,2\}}$, Bob's secret bit $r$, \item[] \hspace{2cm}public threshold $\mathcal{T}$, public field $\mathbb{F}_\text{FF}$ \State $a = [a]^1 + [a]^2$ \If{$a \geq \text{FF}$} \State $b \leftarrow a - $FF \Else \State $b \leftarrow a$ \EndIf \If{$b < \mathcal{T}$} \State $z \leftarrow 1$ \Else \State $z \leftarrow 0$ \EndIf \If{Party = Alice} \State \Return $z \oplus r$ \Else \State \Return $r$ \EndIf \end{algorithmic} \end{algorithm}

\begin{algorithm}
    \caption{$\Pi_\mathsf{VT^-}^\textit{Insert}$} \label{app:insert-}
    \begin{algorithmic}
        \State Input: Share of submitted record $[v]^{\{1,2\}}$, $\mathfrak{D}^{\{1,2\}}$
        \State Output: Updated dataset $\mathfrak{D}^{\{1,2\}}$ per server
        \State $\mathfrak{D}^{\{1,2\}} \leftarrow \mathfrak{D}^{\{1,2\}} \cup [v]^{\{1,2\}}$
        \State \Return
    \end{algorithmic}
\end{algorithm}

\subsection{$\Sigma_\mathsf{VT}$}  \label{app:VT}
The main variant of the scheme follows the same pattern as $\Sigma_\mathsf{VT^-}$, reusing exactly the same GC, B2A, and Data-Prep protocols.

\begin{algorithm*}
    \caption{$\Pi_\mathsf{Intersect}$}
    \begin{algorithmic}[1]
        \State Input: data shares per server $[c]^{\{1,2\}}$, attribute bitlengths $|t|_1, \dots |t_m|$, query bitlength $|t|$, pos$_k$, the index where subvector $k$ starts, $[q]^{\{1,2\}}$, choice of algorithm $\mathfrak{A}$, bank of Beaver triples $[T]^{\{1,2\}}$ and $n$, the number of records 
        \State Output: share of equality test results $[z]^{\{1,2\}}_i$ per server per record
        \If{$\mathfrak{A} = $"Couteau"}
            \State Employ Private ET scheme detailed in [7]
        \ElsIf{$\mathfrak{A} = $"Hybrid"}
            \State Employ Private ET scheme detailed in [7] but replace `product-sharing' function with $\mathfrak{A} = $"intersect2 on the size-reduced strings, 
        \Else
            \State $l \leftarrow |t|$
            \State $l_k \leftarrow |t|_k, 1 \leq k \leq m$
            \State $j_1, j_1^\prime \leftarrow 1$
            \State $\mathcal{S}^{\{1,2\}} \leftarrow \emptyset$ , let $\mathcal{S}$ behave as a stack
            \State $r \leftarrow 0$
            \While{$l > 1$}
                \State $j_1^\prime \leftarrow j_1$
                \State $j_2, j_2^\prime \leftarrow 1$
                \For{$k \in \{1 \dots m\}$}
                    \If{$l_k = 1$}
                    \Comment{if a subvector has been fully evaluated, bring it to the next stage}
                        \For{all $\mathcal{R}_i \in \mathfrak{D}$}
                            \State Push $[c]^1_{i, pos_k} \oplus [q]^1_k$ to stack $\mathcal{S}^1$
                            \State Push $[c]^2_{i, pos_k} \oplus \neg [q]^2_k$ to stack $\mathcal{S}^2$
                            \State $l_k \leftarrow 0$
                        \EndFor
                    \Else
                        \State $P_k \leftarrow \lfloor l_k / 2 \rfloor$
                        \Comment{$P_k$, the number of pairs}
                        \State $l_k \leftarrow l_k - P_k$
                        \State $l \leftarrow l - P_k$
                        \State $s_k \leftarrow 2^r$
                        \Comment{$s_k$, the spacing between the bits}
                        \For{$p_k \in \{1\ldots P_k\}$}
                            \State $i_{k,1} \leftarrow 2p_k \cdot s_k$
                            \State $i_{k,2} \leftarrow s_k \cdot (2p_k + 1)$
                            \For{all $\mathcal{R}_i \in \mathfrak{D}$}  \Comment{multiplication setup on stage 1 bits}
                                \State $[D]^1_{i,(pos_k + p_k)} \leftarrow [c]^1_{i,(pos_k + i_{k, 1})} \oplus [T]^1_{j_1,1}$
                                \State $[E]^1_{i,(pos_k + p_k)} \leftarrow [c]^1_{i,(pos_k + i_{k,2})} \oplus [T]^1_{j_1,2}$
                                \State $[D]^2_{i,(pos_k + p_k)} \leftarrow [c]^2_{i,(pos_k + i_{k,1})} \oplus [T]^2_{j_1,1}$
                                \State $[E]^2_{i,(pos_k + p_k)} \leftarrow [c]^2_{i,(pos_k + i_{k,2})} \oplus [T]^2_{j_1,2}$
                                \State $j_1 \leftarrow j_1 +1$
                            \EndFor
                        \EndFor
                    \EndIf
                \EndFor
                \While{$|\mathcal{S}^1| \geq 2n$}
                        \For{all $\mathcal{R}_i \in \mathfrak{D}$}
                        \Comment{multiplication setup on stage 2 bits i.e. been XORed with q}
                            \State $[D]^1_{i,(l+j_2)} \leftarrow \text{pop}(\mathcal{S}^1) \oplus [T]^1_{j_1,1}$
                            \State $[E]^1_{i,(l + j_2)} \leftarrow \text{pop}(\mathcal{S}^1) \oplus [T]^1_{j_1,2}$
                            \State $[D]^2_{i,(l + j_2)} \leftarrow \text{pop}(\mathcal{S}^2)  \oplus [T]^2_{j_1,1}$
                            \State $[E]^2_{i,(l + j_2)} \leftarrow \text{pop}(\mathcal{S}^2)  \oplus [T]^2_{j_1,2}$
                            \State $j_1 \leftarrow j_1 + 1$
                            \State $j_2 \leftarrow j_2 + 1$
                        \EndFor
                    \EndWhile
                \algstore{alg}
    \end{algorithmic}
\end{algorithm*}

\begin{algorithm*}
    \begin{algorithmic}[1]
        \algrestore{alg}
                \State $j_1 \leftarrow j_1^\prime$
                \State $j_2^\prime \leftarrow j_2$
                \State $j_2 \leftarrow 1$
                \State Servers publish D and E shares
                \For{$k \in \{1 \dots m\}$}
                    \For{$p_k \in \{1\ldots P_k\}$}
                        \State $i_{k,1} \leftarrow 2p_k \cdot s_k$
                        \State $i_{k,2} \leftarrow s_k \cdot (2p_k + 1)$
                        \For{all $\mathcal{R}_i \in \mathfrak{D}$}
                        \Comment{multiplying stage 1 bits}
                            \State $D_{i,(pos_k + p_k)} \leftarrow [D]^1_{i,(pos_k + p_k)} \oplus [D]^2_{i,(pos_k + p_k)}$
                            \State $E_{i,(pos_k + p_k)} \leftarrow [E]^1_{i,(pos_k + p_k)} \oplus [E]^2_{i,(pos_k + p_k)}$
                            \State $[c]^1_{i,i_{k,1}} \leftarrow D_{i,(pos_k + p_k)}\cdot [T]^1_{j,2} \oplus E_{i,(pos_k + p_k)}\cdot [T]^1_{j,1} \oplus [T]^1_{j, 3}$
                            \State $[c]^2_{i,{k,i_1}} \leftarrow D_{i,(pos_k + p_k)}\cdot E_{i,(pos_k + p_k)} \oplus D_{i,(pos_k + p_k)}\cdot [T]^2_{j,2} \oplus E_{i,(pos_k + p_k)}\cdot [T]^2_{j,1} \oplus [T]^2_{j,3}$
                            \State $j_1 \leftarrow j_1 + 1$
                        \EndFor
                    \EndFor
                \EndFor
                \For{$j_2 \in \{1\dots j_2^\prime\}$}
                    \For{all $\mathcal{R}_i \in \mathfrak{D}$}
                    \Comment{multiplying stage 2 bits}
                            \State $D_{i,(l+j_2)} \leftarrow [D]^1_{i,(l+j_2)} \oplus [D]^2_{i,(l+j_2)}$
                            \State $E_{i,(l+j_2)} \leftarrow [E]^1_{i,(l+j_2)} \oplus [E]^2_{i,(l+j_2)}$
                            \State Push $D_{i,(l+j_2))}\cdot [T]^1_{j,2} \oplus E_{i,(l+j_2)}\cdot [T]^1_{j,1} \oplus [T]^1_{j, 3}$ to $\mathcal{S}^1$
                            \State Push $D_{i,(l+j_2)}\cdot E_{i,(l+j_2)} \oplus D_{i,(l+j_2)}\cdot [T]^2_{j,2} \oplus E_{i,(l+j_2)}\cdot [T]^2_{j,1} \oplus [T]^2_{j,3}$ to $\mathcal{S}^2$
                            \State $j_1 \leftarrow j_1 + 1$
                        \EndFor
                    \EndFor
                    \State $r \leftarrow r + 1$
            \EndWhile
            \State \Return $[z]^{\{1,2\}} \leftarrow \mathcal{S}^{\{1,2\}}$
        \EndIf
    \end{algorithmic}
\end{algorithm*}

\begin{algorithm*}
    \caption{$\Pi_\mathsf{VT}^\textit{Insert}$} \label{app:insert}
    \begin{algorithmic}[1]
        \State Input: Share of submitted record $[v]^{\{1,2\}}$, $\mathfrak{D}^{\{1,2\}}$
        \State Output: Updated dataset $\mathfrak{D}^{\{1,2\}}$ per server if and only if no subvector contains the vector of all 1s
        \For{$k \in \{1 \dots m\}$}
            \State Compute $\Pi_\mathsf{Intersect}$ but with subvector $[v]^{\{1,2\}}_k$ instead of $\mathfrak{D}$. Set $q \leftarrow 1$ and $t \leftarrow \{1\}^{l_k}$.
            \State Servers publish $z$ values
            \If{z = 1}
                \State \Return
            \EndIf
        \EndFor
        \State $\mathfrak{D}^{\{1,2\}} \leftarrow \mathfrak{D}^{\{1,2\}} \cup [v]^{\{1,2\}}$
        \State \Return
    \end{algorithmic}
\end{algorithm*}

\section{Beaver Triples \cite{beaver}} \label{triples}
\subsection{Beaver-triple Generation} Beaver triples can be generated with oblivious transfers. To generate a triple in $\mathbb{F}_2^3$ with two servers, both servers $i$ start by randomly sampling $([a]^i, [b]^i) \in \mathbb{F}_2^2$. The servers wish to calculate $[c]^i$ such that $c^1 + c^2 = c$ and $([a]^1 + [a]^2) \cdot ([b]^1 + [b]^2) = c$. Server 1 chooses a random bit $r_1$ and acts as the sender of an OT with input $(r_1, r_1 \oplus [a]^1)$. Server 2 selects with $[b]^2$, and thus learns $[a]^1 [b]^2 \oplus r_1$. This process is then repeated in the opposite direction, such that server $R$ learns $[a]^2 [b]^1 \oplus r_2$. Server 1 calculates $[c]^1 = r_1 \oplus [a]^1 [b]^1 \oplus [a]^2 [b]^1 \oplus r_R$ and server 2 calculates $c_2 = r_2 \oplus [a]^2 [b]^2 \oplus [a]^1 [b]^2 \oplus r_1$, and it is easily seen that:
\vspace{10pt}

\noindent
$[c]^1 \oplus [c]^2 = r_1 \oplus r_1 \oplus r_2 \oplus r_2 \oplus [a]^1 [b]^1 \oplus [a]^2 [b]^2 \oplus [a]^1 [b]^2 \oplus [a]^2 [b]^1$
\newline \text{} \hspace{38pt} $= [a]^1 [b]^1 \oplus [a]^2 [b]^2 \oplus [a]^1 [b]^2 \oplus [a]^2 [b]^1$
\vspace{10pt}

Recalling that XOR-ing is equivalent to addition in $\mathbb{F}_2$:

\vspace{10pt}
\noindent
$[c]^1 + [c]^2 = [a]^1 [b]^1 + [a]^2 [b]^2 + [a]^1 [b]^2 + [a]^2 [b]^1$

\vspace{5pt}
\text{} \hspace{27pt} $= ([a]^1 + [a]^2) \cdot ([b]^1 + [b]^2)$

\vspace{10pt}
and hence $c = a \cdot b$. It is also possible to extend these triples to larger fields by using \emph{oblivious linear evaluation}.

\begin{figure*}
    \centering
    \includegraphics[width=0.8\linewidth]{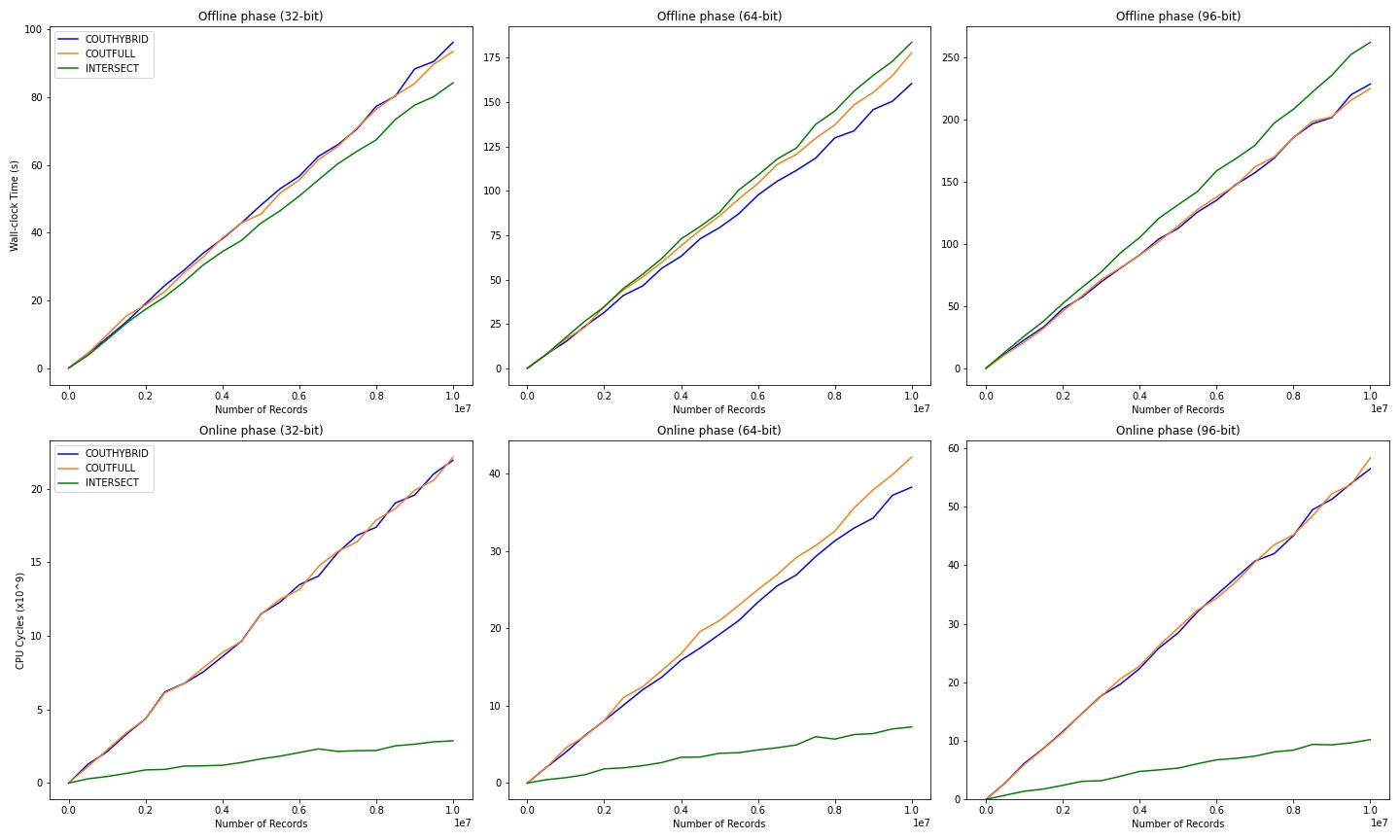}
    \caption{Additional benchmarks showing online and offline phases in alternative metrics}
    \label{fig:plots3}
\end{figure*}
\subsection{Beaver-triple Evaluation}

 Each server $i$ starts with $[x]^i$, $[y]^i$ and $([a]^i, [b]^i, [c]^i)$, and wishes to calculate $[z]^i$ such that $z = x \cdot y$. Server $i$ begins by calculating $[d]^i = [x]^i - [a]^i$ and $[e]^i = [y]^i - [b]^i$. The servers publish their shares of $d$ and $e$ such that they are known in the clear, which is safe information to share, due to the randomness at which the Beaver triple was calculated. Then, $[z]^i$ is evaluated as $\frac{1}{2}(d \cdot e) + d \cdot [b]^i + e \cdot [a]^i + [c]^i$. Correctness is easily verified: 

\noindent$\sum_i (\frac{1}{2}(d \cdot e) + d \cdot [b]^i + e \cdot [a]^i + [c]^i)$
\newline
$= d \cdot e + d \cdot b + e \cdot a + c$
\newline
$= (x - a) \cdot (y - b) + (x - a) \cdot b + (y - b) \cdot a + c$
\newline
$= (x - a) \cdot y + (y - b) \cdot a + c$
\newline
$= x \cdot y - a \cdot y + a \cdot y - a \cdot b + c$
\newline
$= x \cdot y - a \cdot b + c$
\newline
$= x \cdot y$ \hspace{20pt} 

As an intuition, this approach relies on the random Beaver triple to hide the cross products $[x]^1 \cdot [y]^2$ and $[x]^2 \cdot [y]^1$, meanwhile the triple itself is eliminated in the calculation of $z$.

\section{B2A Conversion \cite{prio+}} \label{B2A}
\subsubsection*{daBit Generation}
To generate a daBit share $([b]^A, [b]^B)$ where $[b]^B$ is a Boolean share and $[b]^A$ is an arithmetic share in the field $\mathbb{Z}_p$, the servers perform the following protocol, taken directly from \cite{prio+}:
\begin{enumerate}
    \item Both servers sample a random bit $b_i \in {0, 1}$. Server 1 samples a random integer $x \in \mathbb{F}_p$
    \item Server 1 acts as the OT sender, sending $(x, x + b_1)$, and sets $y_1 = -x$mod $p$
    \item Server 2 acts as the OT receiver, using $b_2$ as the choice bit, receiving $y_2 = x + b_1 b_2$ mod $p$
    \item Both servers set $[b]^A_i = b_i - 2y_i$mod $p$ and $[b]^B_i = b_i$.
\end{enumerate}

\subsubsection*{daBit Consumption}
To convert a bit $x$ to an arithmetic share, the servers perform the following protocol:
\begin{enumerate}
    \item Both servers compute $[v]^B_i = [x]^B_i \oplus [v]^B_i$
    \item Both servers publish their shares $[v]^B_i$
    \item Both servers calculate $v$ as $v = x \oplus b = [v]^B_1 \oplus [v]^B_R$
    \item Server 1 outputs $[x]^A_1 = v + [b]^B_1 - 2v[b]^B_1$mod$p$, and server 2 outputs  $[x]^A_2 = [b]^B_2 - 2v[b]^B_2$mod$p$.
\end{enumerate}

\section{Further Benchmarks} \label{app:furtherbenchmarks}
We include benchmarks for offline phase in wall=clock time and online phase in CPU cycles for the reader who wishes to compare the cost of the phases in the same metric. This can be seen in figure \Cref{fig:plots3}. We stress that the metrics for the corresponding phase is not ideal --- in solidarity, it paints the online phase in a better light than it should be due to the disregard of communication time and the offline phaes in a worse light since it has not been parallelised and the clock speed used in the benchmark is relatively low. 

\section{Round Trip Derivations} \label{rtt}
\subsection{$\mathsf{\Pi}_{{Intersect}^-}$}
Our first problem is to consider how many round trips are required for an algorithm optimised for round trips on the protocol $\mathsf{\Pi}_{{Intersect}^-}$. When performing the real protocol, communication is required every time we evaluate a multiplication. Fortunately, we can do this in parallel. It is thus sufficient to say that a round trip (henceforth, a 'round') is required every time the input of the distributed multiplication function (2 fan-in, 1 fan-out) is an output of the function from the same round. 

The problem can now be viewed as a recursive size-reducing protocol, where at every round, we take as many pairs of bits as possible and input them to the function (which can now by regarded as a black box) such that we get 1 bit output for every pair. This clearly forms a binary tree, where we merge nodes as early as possible. Thus, it intuitively follows that the number of round trips for $\mathsf{\Pi}_{{Intersect}^-}$ is $R_1^{*} = \lceil \text{log}_2l\rceil$.

\subsection{$\mathsf{\Pi}_{Intersect}$}
We now add the constraint that a subvector is to be fully evaluated (only 1 bit remains) before it can be input into the function. A naive approach is to evaluate all subvectors (max$_i\lceil \text{log}_2l_i \rceil$ rounds) before combining the remaining bits (1 bit per attribute, $\lceil \text{log}_2m \rceil$ rounds for $m$ attributes), to get $R_2 = \text{max}_i\lceil \text{log}_2l_i \rceil + \lceil \text{log}_2m \rceil$. An optimal scheduler can, however, do better by evaluating pairs that are ready (from fully evaluated subvectors) while larger subvectors are still merging. 

To tackle this problem, we distinguish 2 different 'views'. $\mathcal{V}_1$ is the straightforward view of the binary tree that merges inputs until only the root remains. $\mathcal{V}_2$ is then an auxiliary binary tree used to analyse the optimal round count. In the new view, the leaves of the tree are the number of rounds required to fully evaluate each subvector, $d_i = \lceil \text{log}_2l_i \rceil$. We continue building the new view binary tree by merging nodes according to the function max$(a, b) + 1$ for parents $a$ and $b$. Our optimisation problem is the min-max problem of minimising max$_i(d_i + h_i)$, where $h_i$ is the distance from the node to the root. If we require that each leaf node be at the lowest depth before it is merged, $\mathcal{V}_2$ enumerates a prefix code (since no node has exactly one child), and thus leverages the Kraft-McMillan theorem which asserts that a uniquely decodable code and a prefix code exists for a given set of codeword lengths if and only if their lengths satisfy $\sum_{i=1}^m2^{-h_i} \leq 1$, for a binary alphabet and codeword lengths $h_i$. We arrive at the optimal round count using the following derivations:

\begin{enumerate}
    \item Let $D$ be the depth of the tree in $\mathcal{V}_1$ and the codeword length $h_i$ is the distance between leaf $d_i$ and the root of the tree. It follows that $D = \text{max}_i(d_i + h_i)$, and thus $D \geq d_i + h_i$ for any $i$. We use the Kraft inequality to arrive at the lower bound. Letting $S=\sum_{i=1}^m2^{d_i}$, we have $$1 \geq \sum_{i=1}^m2^{-h_i} \geq \sum_{i=1}^m2^{-(D-d_i)} = 2^{-D}\sum_{i=1}^m2^{d_i}=2^{-D}S$$ which shows that any valid $D$ satisfies $D\geq \text{log}_2S$. 
    \item Since every $D$ follows this bound, and since $D$ and the optimal distance $D^{*}$ are integers, it follows that $D^{*} \geq \lceil \text{log}_2S\rceil$.
    \item The Kraft-McMillan theorem asserts that, given a set of non-negative integers $h_i$ satisfying the Kraft inequality, there exists a feasible tree with depths $h_i$. We propose that a tree exists that has leaves $h_i = \lceil \text{log}_2S \rceil - d_i$, for all possible $i$. The leaves are guaranteed to be non-negative since $2^{d_i} \leq S \implies d_i \leq \text{log}_2S \implies d_i \leq \lceil \text{log}_2S \rceil \implies h_i =  \lceil \text{log}_2S \rceil - d_i \geq 0$. Checking the Kraft inequality, we have $\sum_{i=1}^m2^{-h_i} = 2^{-\lceil \text{log}_2S \rceil}S \leq 1$, which holds because $\lceil \text{log}_2S \rceil \geq \text{log}_2S$. Therefore, the optimal distance must follow $D^{*} \leq \lceil \text{log}_2S \rceil$.
    \item Since $D^{*} \leq \lceil \text{log}_2S \rceil$ and $D^{*} \geq \lceil \text{log}_2S \rceil$, it follows that $D^{*} = \lceil \text{log}_2S \rceil$.
\end{enumerate}

We now evaluate $R_2^{*}$. Firstly, every scheduler has $D \geq D^{*}$, therefore $R_2^{*} \geq D^{*}$. Secondly, the explicit construction $h_i = D^{*}-d_i$ is a valid schedule since it satisfies the Kraft-McMillan theorem, and the finishing round is exactly $D^{*}$. It follows that $R_2^{*} \leq D^{*}$. Since $R_2^{*} \leq D^{*}$ and $R_2^{*} \geq D^{*}$, it follows that $R_2^{*} = D^{*}$, and finally, $R_2^{*} = \lceil \text{log}_2(\sum_{i=1}^m2^{\lceil \text{log}_2l_i\rceil})\rceil$.

We informally argue that the optimal scheduler greedily merges nodes with the lowest 2 values - intuitively, merging with higher values can only perform worse, since the combining function takes the maximum.

\subsection{Bound of Additional Cost}
We now prove that $R_1^{*} \leq R_2^{*} \leq R_1^{*} + 1$. Since $d_i = \lceil \text{log}_2l_i \rceil$, $d_i$ is the smallest integer such that $2^{d_i} \geq l_i$. It also follows that $d_i-1 < \text{log}_2l_i \implies 2^{d_i - 1} <l_i \implies 2^{d_i} < 2l_i$. Combining this with $l_i \leq 2^{d_i}$ entails $l_i \leq 2^{d_i} < 2l_i$. Summing over all $m$ subvectors, we get $l = \sum_{i=1}^ml_i \leq S = \sum_{i=1}^m2^{d_i} < 2 \sum_{i=1}^ml_i = 2l$. Since $R_1^{*} = \lceil \text{log}_2l \rceil$, it follows that $2^{R_1^{*}-1}<l\leq 2^{R_1^{*}}$.

\begin{enumerate}
    \item Since $S \geq l > 2^{R_1^{*}-1}$, it follows that log$_2S>R_1^{*} -1$, and thus $R_2^{*} = \lceil \text{log}_2S \rceil \geq R_1^{*}$.
    \item Since $S <2l \leq 2^{R_1^{*}+1}$, it follows that log$_2S<R_1^{*} +1$, and thus $R_2^{*} = \lceil \text{log}_2S \rceil \leq R_1^{*} + 1$.
\end{enumerate}
We therefore have that $R_1^{*} \leq R_2^{*} \leq R_1^{*} + 1$.

\section{Data Poisoning Attacks}
Our correctness, robustness and security properties demonstrate that our system correctly computes the threshold function with the data provided, does not reveal additional information than the query itself, and does not permit users to impact the system more than misreporting their own data. Although the scoping of these properties is in-line with the norm for similar works, the properties should not be considered as all-encompassing. For example, our system guarantees that not more information than the result of the query is leaked, but does not guarantee that the output itself retains privacy in a more generic sense. In this section, we discuss some of the avenues of attack that are not (and should not be) in scope of the above properties. We consider a data poisoning attack to be successful if it causes a user's population to be deemed safe when it is not so in the real population (population excluding poisoned data). 

\subsection{Junk Poisoning}
One trivial attack is to flood the database with random data. As with any data aggregation, the result is only as useful as the data itself. As the dataset gets flooded with junk, a given population with low count will approach the count expected from a random dataset, which may push the count to exceed the threshold. The probability that a randomly generated record increments the count of a query is mathematically negligible in the number of attributes. Moreover, the count needs to increment up to $\mathcal{T}$ times, while the threshold should scale with the size of the dataset. Even for small queries, an attacker must launch a large number of queries to succeed in their attack. We leave mitigation of this attack to the software level, which may be done through traffic analysis.

\subsection{User Attack} 
We consider an adversary who wishes to cause a targetted user to erroneously be told that their population is safe. We observe that do this effectively, the attacker must already know the populations to which the user belongs. An adversary with this information can succeed in their attack by sending many duplicate queries. This can be mitigated as before with packet analysis. Additionally, an implementation could easily test whether several entries are equal to each other by accumulating a safe amount of entries (say, 100) and performing checks to test for equality. The servers can drop the buffer if there is a suspicious number of equal entries. To test for equalities, the servers go through each entry in the buffer and count how many entries are equal to that entry by using the record itself as the test vector and the threshold $\mathcal{T}$ sensibly chosen. 

Despite these and other possible mitigations, it would be very difficult to entirely stop this type of attack without requiring authentication, which would be too inconvenient for users in the targetted use case. Any data the adversary does not know increases the complexity exponentially, as with junk poisoning. Since our system is intended to be an informative tool for moderately concerned individuals (rather than high-level targets) then an attacker does not have much to gain and we do not expect a high complexity from the adversary. Additionally, the requirement to know a large part of the victim's data reduces the number of viable attackers further. As such, we choose to accept this vulnerability. 

\subsection{Attribute Attack} We consider an adversary that wishes to corrupt a specific attribute/value. For a targetted value attack, We repeat the same argument as for junk poisoning, except the adversary uses a constant value for one attribute which reduces the complexity exponent by 1 (still exponential). For a targetted attribute attack, the attacker must use every combination of values, which is equivalent to the junk poisoning attack.

\section{Probabilistic Optimisation with Parity Samples}
We can adjust our protocol such that its computational complexity is negligible in the bitlength. In order to do so, we must move to a probabilistic scheme. More specifically, we use random parity samples to give us a scheme that achieves arbitrary accuracy or bitlength complexity. A complexity of O($n\ell$) gives a $\frac{1}{2^\ell}$ false positive rate (with  0\% false negative and perfect soundness).

The scheme works by calculating the distributed OR over the output of $\ell$ parity checks. For each parity check, the servers check the parity of a randomly generated list of indices, which can be publicly generated from a PRNG. If the two strings are the same, the parity will always match. If not, there is a 50\% chance that the parity matches. If the servers perform many samples, they can compute the distributed OR function over the outputs and conclude that the strings match with $1-\frac{1}{2^\ell}$ confidence if all parity checks are equal or conclude they do not match if any of the checks are not equal. Distributed OR can be performed simply by negating the inputs and outputs of the distributed AND function.

To highlight its effectiveness, consider the following example. For a query of very large bitlength (say, 2048), we could accept a false positive rate of $2^{-32}$ to drastically reduce the computational complexity. The servers must perform many local XORs, but these are cheap and the benefit is that the servers would expend $\frac{32}{2048}$x as much pre-computation (Beaver triples). For a dataset of 1m clients where 10 of the records actually match, the probabilistic scheme would output $10+\frac{1000000}{2^{32}} \approx 10.00023$ matches on average.

\section{Query Expressiveness} 
Our protocol can efficiently answer a wide range of queries. We informally argue functional completeness for our queries. In particular, functional completeness means that our main protocol $\mathsf{\Pi}_{Intersect}$ can answer a query that is built using any combination of AND, OR, or NOT. In its simplest form, a query is a finite sequence of attribute constraints in conjunctive normal form, using only OR, AND and NOT. Any bracketed parts of the expression can be brought to their simplified equivalent using basic logic algebra, e.g. $\{\neg (\mathcal{A}_1 = A_1 \lor \mathcal{A}_1 = A_2)\} \iff \{(\mathcal{A}_1 \neq A_1 \land \mathcal{A}_1 \neq A_2)\}$.

In general, each OR costs an additional query to the dataset, while AND and NOT are free.

\subsection{AND}
The AND operator between two values within different attributes is the main focus in this paper and has been described at length. In particular, we show proof of correctness in \Cref{correctness}. By construction, an attribute cannot have more than one value, so querying AND between two values within the same attribute is nonsensical.

\subsection{NOT}
The clever construction of $q$ already enables a NOT operator. The value of $\neg q$ dictates whether to negate the result of the equality test for a certain attribute, i.e. it tests whether the test vector $t$ does \emph{not} equal the corresponding subvector in the dataset. We thus allow the requesting client to arbitrarily `disregard' attributes by setting $\neg q$ to 1 and $t$ to the vector of all 1s $\{111\dots \}$. This computes exactly NOT($t$), which is equivalent to the union of all other values in that attribute. Since no attribute has the vector of all 1s, this always returns true from the equality test, enabling the main functionality of the protocol. A client can then achieve arbitrary NOT by simply choosing $t$ and letting $\neg q = 1$.

\subsection{OR}
A straightforward way to achieve OR between two values within the same attribute is to simply add the count of two queries together. For example, we could evaluate count($\mathcal{Q}) = \text{count}(\{(\mathcal{A}_1 = A_1 \lor \mathcal{A}_1 = A_2) \land \Gamma\})$ as count($\{\mathcal{A}_1 = A_1 \land \Gamma\}) + $ count($\{\mathcal{A}_1 = A_2 \land \Gamma\})$ for attribute $\mathcal{A}_1$, values $A_1, A_2 \in \mathcal{A}_1$, and the arbitrary remaining part of the query $\Gamma$. This requires computing over all the data twice (+100\% overhead). Note that the count shares are summed locally before the GC. The finite field size must be doubled, but the GC remains the same. 

Note that a much faster construction exists for $\mathsf{\Pi}_{{Intersect}^-}$, which requires leaking which attributes contain an OR operator in the query. The construction entails performing the ET over every value in the conjunction and simply summing locally the shares of the results. Since a maximum of only 1 value can `pass', the real sum does not overflow the modulus (2). The resulting construction then has an overhead equal to the cost of computing the additional ETs. 

To compute OR between two values within different attributes (for either protocol variant), the client can take the difference between the query with those attribute disregarded and the query that includes the disjunction of each negated attribute. In other words, count$(\{(\mathcal{A}_1 = A_1 \lor \mathcal{A}_2 = A_2) \land \Gamma\})$ for $A_1 \in \mathcal{A}_1, A_2 \in \mathcal{A}_2$, can be computed as count$(\{\Gamma\}) - $ count$(\{\mathcal{A}_1 \neq A_1 \land \mathcal{A}_2 \neq A_2) \land \Gamma\})$. Again, this has a +100\% overhead.

\section{Security of building blocks} \label{app:ideals}
\paragraph*{Ideal Functionality}

\begin{definition}
    ($\mathfrak{I}_{\text{Mult}}$). The two-party ideal functionality $\mathfrak{I}_{\text{Mult}}$ behaves as follows:
    \begin{itemize}
        \item $\mathfrak{I}_{\text{Mult}}$ receives secret shares $([x]^i, [y]^i) \in \mathbb{F}_2^2$ from server $i$
        \item $\mathfrak{I}_{\text{Mult}}$ computes $z = xy$
        \item $\mathfrak{I}_{\text{Mult}}$ returns a fresh random sharing $[z]^i$ to server $i$
    \end{itemize}
\end{definition}

\begin{definition}
    ($\mathfrak{I}_{\text{B2A}}$). The two-party ideal functionality for $\mathfrak{I}_{\text{B2A}}$ for Boolean inputs and outputs in the public field $\mathbb{F}_{FF}$ behaves as follows:
    \begin{itemize}
        \item $\mathfrak{I}_{\text{B2A}}$ receives secret shares $[x]^i \in \mathbb{F}_2$ and pre-computation shares $([b]^i, [a]^i) \in \mathbb{F}_2 \times \mathbb{F}_{FF}$ from server $i$
        \item $\mathfrak{I}_{\text{B2A}}$ computes $z = x$ where $z \in \mathbb{F}_{FF}$
        \item $\mathfrak{I}_{\text{B2A}}$ returns $[z]^i$ to server $i$ (where $[z]^i$ is an arithmetic share of $z$)
    \end{itemize}
\end{definition}

\begin{definition}
    ($\mathfrak{I}_{\text{GC}}$). The two-party ideal functionality $\mathfrak{I}_{\text{GC}}$ \newline parametrised by public field size FF and public threshold $\mathcal{T}$ behaves as follows:
    \begin{itemize}
        \item $\mathfrak{I}_{\text{GC}}$ receives $[Z]^1 \in \mathbb{F}_{FF}$ from $S_1$ and $[Z]^2 \in \mathbb{F}_{FF}$ from $S_2$
        \item $\mathfrak{I}_{\text{GC}}$ computes $Z = [Z]^1 + [Z]^2$ mod $FF$ (as per the circuit defined in \Cref{app:gc})
        \item $\mathfrak{I}_{\text{GC}}$ returns the bit $b =(Z < \mathcal{T})$ to the garbler, server 1
    \end{itemize}
\end{definition}
\begin{lemma}[\textbf{Sequential Composition}, semi-honest \cite{composition}]
\label{lem:composition}
Let $\pi$ be a protocol that invokes ideal functionalities $\mathfrak{I}_1, \dots, \mathfrak{I}_k$ as subroutines and suppose $\pi$ securely computes $f$ in the $(\mathfrak{I}_1, \dots ,\mathfrak{I}_k)$-hybrid model against the adversary structure of \Cref{def:react-f-privacy}. Suppose further
that each $\mathfrak{I}_j$ is securely realized, against the same adversary structure, by a real-world sub-protocol $\rho_j$. Then the composed protocol $\pi^{\rho_1,\dots,\rho_k}$, obtained by replacing each ideal call to $\mathfrak{I}_j$ with an execution of $\rho_j$, securely computes $f$ in the real model.
\end{lemma}

\section{Example Use Case} \label{app:usecase}
Suppose we have a user who wishes to query whether the population of people who are heterosexual women aged 30-40 is vulnerable (i.e. whether the count exceeds the threshold) and creates the query $\{1110||01|0100|01\}$ as before. The client will choose 2 random shares that XOR to $01|0100|01$, e.g. $[t]^1 = 01|0101|11), [t]^2 = 00|0001|10$ and sends each to the corresponding server. Server 2 additionally negates $[t]^2$, and the servers apply the bitmask to each record in the database (taking only the relevant attributes). If we suppose there is a record in the database whose data in the clear $[v]$ would be the same as $t$
but with different shares, e.g. $[v]^1 = 00|1001|11$ and $[v]^2 = 01|1101|10$ we can apply the bitmasks $[t]^1$ and $\neg[t]^2$ to obtain $[v]^1 \oplus [t]^1 = 01|1100|00$ and $[v]^2 \oplus \neg[t]^2 = 10|0011|11$ and observe that the recombination of these values is the vector of all 1s, i.e. $([v]^1 \oplus [t]^1) \oplus ([v]^2 \oplus \neg[t]^2) = 01|1100|00 \oplus 10|0011|11 = 11|1111|11$, and thus, clearly,  $([v]^1 \oplus [t]^1) \oplus ([v]^2 \oplus \neg[t]^2)$ equals the vector of all 1s if and only if $v = t$. After masking the data, the servers then utilise the secret multiplication protocol to compute the AND function of the vector 11|1111|11 (which is not visible in the clear), which should output 1, as we have a match. In more detail, the normal multiplication function *$(\cdot, \cdot)$ that takes 2 parameters is related to a distributed multiplication protocol $dm$ in the following way: $*(x, y) = z = dm^1([x]^1, [y]^1) \oplus dm^2([x]^2, [y]^2) = [z]^1 \oplus [z]^2$.

Therefore, to test whether $(01|1100|00) \oplus (10|0011|11) = $ \newline$(11|1111|11)$, we perform the following calculation (simplified to the first 4 bits):
$$ 1*1*1*1= *(*(*(1, 1), 1), 1)$$
$$dm^1_1(dm^1_2(dm^1_3(0, 1), 1), 1) \oplus dm^2_1(dm^2_2(dm^2_3(1, 0), 0), 0)$$

The equality test requires the product of all $\ell$ masked bits. These multiplications can be arranged as either a sequential chain or a balanced tree. The former minimises the number of intermediate values but requires $O(\ell)$ sequential communication rounds, whereas the latter evaluates independent products in parallel and requires only $O(\log_2\ell)$ rounds. This choice illustrates the trade-off; we use the balanced ordering in our implementation to minimise online latency.

\section{Open Science}
The implemented code can be found publicly at \url{https://anonymous.4open.science/r/privateintersect-F711/}. A list of dependencies and installation instructions can be found in the readme.

\section{Ethics}
This work does not involve human subjects, collection of personal data, or experiments involving individual users. All date used to evaluate our protocol was synthetically created.

The proposed system is intended to reduce privacy risks arising from the combination of attributes across a population by allowing threshold queries to be evaluated without revealing individual records or the queried attributes to the servers. The capability could nevertheless be used to investigate sensitive populations, and therefore has potential for misuse. Our construction does not eliminate this risk: in particular, a querying client learns the threshold outcome for its chosen query and may issue multiple queries. We therefore explicitly model the information revealed by such queries through the ideal functionality and reactive $f_T$-privacy. Practical deployments should additionally consider appropriate authorisation, query-rate limits, and governance to prevent systematic abuse of the functionality.
\end{document}